\documentstyle[11pt,amssymb,epsfig,psfig]{article}

\makeatletter

\@addtoreset{equation}{section}

\makeatother

\begin{document}

\title{On the energy-momentum tensor for a scalar field on
manifolds with boundaries}
\author{Aram A. Saharian\footnote{%
Email: saharyan@server.physdep.r.am}\\
\textit{Department of Physics, Yerevan State
University, 1 Alex Manoogian Str.,}\\
\textit{375049 Yerevan, Armenia,}\\
\textit{and}\\
\textit{the Abdus Salam International Centre for Theoretical
Physics,}\\
\textit{34014 Trieste, Italy }}
\date{\today}
\maketitle
\begin{abstract}
We argue that already at classical level the energy-momentum
tensor for a scalar field on manifolds with boundaries in addition
to the bulk part contains a contribution located on the boundary.
Using the standard variational procedure for the action with the
boundary term, the expression for the surface energy-momentum
tensor is derived for arbitrary bulk and boundary geometries.
Integral conservation laws are investigated. The corresponding
conserved charges are constructed and their relation to the proper
densities is discussed. Further we study the vacuum expectation
value of the energy-momentum tensor in the corresponding quantum
field theory. It is shown that the surface term in the
energy-momentum tensor is essential to obtain the equality between
the vacuum energy, evaluated as the sum of the zero-point energies
for each normal mode of frequency, and the energy derived by the
integration of the corresponding vacuum energy density. As an application,
by using the zeta function technique, we evaluate the surface energy for a quantum scalar field confined inside a spherical shell.
\end{abstract}

\bigskip

PACS numbers: 03.50.Kk, 03.70.+k, 04.62.+v, 11.30.-j

\bigskip

\section{Introduction} \label{sec:introd}

In many problems we need to consider the physical model on
background of manifolds with boundaries on which the dynamical
variables satisfy some prescribed boundary conditions. In presence
of boundaries new degrees of freedom appear, which can essentially
influence the dynamics of the model. The incomplete list of
branches where the boundary effects play an important role
includes: surface and finite-size effects in condensed matter
theory and statistical physics, theory of phase transitions and
critical phenomena \cite{Dieh86,Card87}, canonical formulation of
general relativity (see \cite{Isen80}), the definition of the
gravitational action, Hamiltonian and energy-momentum
\cite{Regg74}--\cite{Brow02}, path-integral approach to quantum
gravity and the problem of boundary conditions for the quantum
state of the universe (see, for instance, \cite{Espo97} and
references therein), bag models of hadrons in QCD, string and
M-theories, various braneworld models, manifolds with horizons and
the thermodynamics of black holes, AdS/CFT correspondence
\cite{Ahar00}, holographic principle \cite{Smol01,Bous02}. In
recent years, the study of the boundary effects in quantum theory
has produced several important results. For example, an
interesting idea is to consider the black hole event horizon as a
physical boundary \cite{Carl00}. This induce an extra term in the
action, having as consequence the existence of a central charge in
the algebra of generators of gauge transformations. Using this
central charge it is possible to determine the asymptotic behavior
of the densities of states and in this way to get the entropy for
a black hole \cite{Card86}. In the context of string theory, the
D-branes are natural boundaries for the open strings, having very
interesting effects in the theory.

In quantum field theory the influence of boundaries on the vacuum
state of a quantized field leads to interesting physical
consequences. The imposition of boundary conditions leads to the
modification of the zero--point fluctuations spectrum and results
in the shifts in the vacuum expectation values of physical
quantities, such as the energy density and vacuum stresses. In
particular, vacuum forces arise acting on constraining boundaries.
This is the familiar Casimir effect. The particular features of
the resulting vacuum forces depend on the nature of the quantum
field, the type of spacetime manifold and its dimensionality, on
the boundary geometries and the specific boundary conditions
imposed on the field. Since the original work by Casimir in 1948
\cite{Casi48} many theoretical and experimental works have been
done on this problem, including various types of bulk and boundary
geometries (see, e.g., \cite{Most97}--\cite{Milt02} and references
therein). The Casimir force has recently been measured with a
largely improved precision \cite{Lamo97} (for a review see
\cite{Bord01,Lamb03}) which allows for an accurate comparision
between the experimental results and theoretical predictions. An
essential point in the investigations of the Casimir effect is the
relation between the mode sum energy, evaluated as the sum of the
zero-point energies for each normal mode of frequency, and the
volume integral of the renormalized energy density. For scalar
fields with general curvature coupling in Ref. \cite{Rome02} it
has been shown that in the discussion of this question for the
Robin parallel plates geometry it is necessary to include in the
energy a surface term concentrated on the boundary (see Refs.
\cite{Saha01,Rome01} for similar issues in spherical and
cylindrical boundary geometries and the discussion in Ref.
\cite{Full03}). However, in Refs. \cite{Rome02,Saha01,Rome01} the
surface energy density is considered only for the case of the
Minkowski bulk with static boundaries. In the present paper, by
using the standard variational procedure, we derive an expression
of the surface energy-momentum tensor for a scalar field with
general curvature coupling parameter in the general case of bulk
and boundary geometries. It is well-known that the imposition of
boundary conditions leads to a vacuum expectation value of the
energy-momentum tensor with non-integrable divergences as the
boundary is approached \cite{Bali78,Deut79,Kenn80}. The role of
surface action to cancel these divergencies is emphasized in Ref.
\cite{Kenn80}, where it has been shown that the finiteness of the
total energy of a quantized field in a bounded region is a direct
consequence of renormalization of bare surface actions. How boundary corrections are implemented by surface interaction and the structure of the corresponding counterterms in the renormalization procedure are discussed in Ref. \cite{Syma81} within the Schr\"odinger representation of quantum field theory. By suitably modifying the classical action with boundary corrections, in Ref. \cite{Vanz93} it has been shown that for scalar fields confined in a cavity familiar functional methods can be applied and the functional measure is constructed by treating the boundary conditions as field equations coming from a variational procedure. Surface
terms play an important role in the recently proposed procedure
\cite{Bala99} (see also \cite{Krau99}--\cite{Noji00} and
references therein) to regularize the divergencies of the
gravitational action on noncompact space. By adding suitable
boundary counterterms to the gravitational action, one can obtain
a well-defined boundary energy-momentum tensor and a finite
Euclidean action for the black hole spacetimes.

This paper is organized as follows. In Sec. \ref{sec:action} the
notations are introduced for the geometrical quantities describing
the manifold and the structure of the action is discussed. In Sec.
\ref{sec:emtbs} the variations of the bulk and surface actions are
considered with respect to variations of the metric tensor and the
expressions for the bulk and surface energy-momentum tensors are
derived. The integral conservation laws described by these tensors
are studied in Sec. \ref{sec:intcons}. The expressions for the
corresponding energy-momenta are derived and relations to the
proper densities are discussed. The vacuum expectation values of
the energy-momentum tensor for a scalar field on a static manifold
with boundary are considered in Sec. \ref{sec:vacemt}. It is
argued that the surface part of the energy density is essential to
obtain the equality between the vacuum energy, evaluated as the
sum of the zero-point energies for each normal mode, and the
energy obtained by the integration of the corresponding vacuum
energy density. As an application of general foemula, in Sec. \ref{sec:sphen} we consider the surface energy for a quantum scalar field confined inside a spherical shell. Section \ref{sec:conc} concludes the main results of
the paper. An integral representation of the partial zeta function needed for the calculation of the surface energy on a spherical shell is derived in Appendix \ref{sec:app1}.

\section{Notations and the action} \label{sec:action}

\begin{figure}[tbph]
\begin{center}
\epsfig{figure=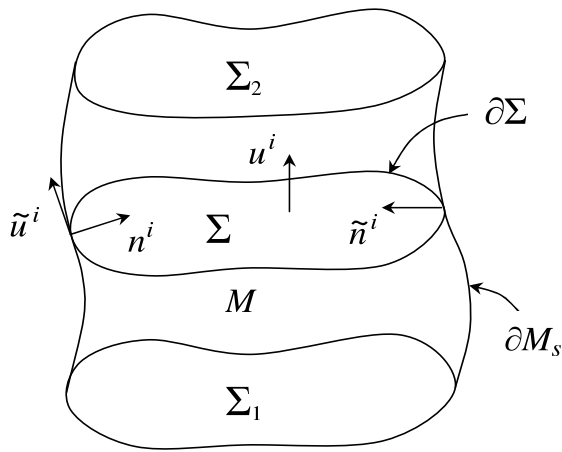,width=8cm,height=6cm}
\end{center}
\caption{Manifold $M$ with boundary $\partial M=\Sigma _1 \cup
\Sigma _2 \cup \partial M_s$ foliated by spacelike hypersurfaces
${\Sigma}$ with boundaries $\partial {\Sigma}$. $n^i$ and $u^i$
are unit normal vector fields to the hypersurfaces $\partial M_s$
and $\Sigma $ respectively. $\tilde n^i$ is the unit normal vector
to the $(D-1)$-surface $\partial \Sigma $ in the hypersurface
$\Sigma $, and $\tilde u^i$ is the unit normal vector to $\partial
\Sigma $ in the hypersurface $\partial M_s$.} \label{fig1fold}
\end{figure}

Consider a $D+1$ - dimensional spacetime region $M$ with metric
$g_{ik}$ and boundary $\partial M$. We assume that the spacetime
$M$ is foliated into a family of spacelike hypersurfaces
${\Sigma}$ defined by $t={\mathrm{const}}$. The boundary of $M$
consists of the initial and final spacelike hypersurfaces $\Sigma
_1 $ and $\Sigma _2$, as well as a timelike smooth boundary
$\partial M_s$ and, hence, $\partial M=\Sigma _1 \cup \Sigma _2
\cup
\partial M_s$. The boundary of each ${\Sigma}$ we will denote $\partial
{\Sigma}$: $\partial \Sigma =\Sigma \cap \partial M_s$. The
corresponding spacetime region is depicted in Fig. \ref{fig1fold}
for a three dimensional $M$. The inward pointing unit normal
vector field on $\partial M$ is denoted $n^i$. This vector is
normalized in accordance with $n_in^i=\epsilon $, where $\epsilon
=1$ and $-1$ for spacelike and timelike boundary elements,
respectively. We define $u^i$ as the future pointing unit normal
vector to the spacelike hypersurfaces $\Sigma $. On the
hypersurfaces $\Sigma _1$ and $\Sigma _2$ this vector is connected
to the vector $n^i$ by relations: $u^i=n^i$ on $\Sigma _1$ and
$u^i=-n^i$ on $\Sigma _2$. In the consideration below we will not
assume that the timelike boundary $\partial M_s$ is orthogonal to
the foliation of the spacetime. We define $\tilde u^i$ as the
forward pointing timelike unit normal vector field to the surfaces
$\partial \Sigma $ in the hypersurface $\partial M_s$. By
construction one has $\tilde u^i n_i=0$. We further define $\tilde
n^i$ as the vector field defined on $\partial M_s$ such that
$\tilde n ^i$ is the unit normal vector to $\partial \Sigma $ in
hypersurface $\Sigma $ and, hence, by construction $u^i \tilde
n_i=0$. It can be seen that the following relations take place
between the introduced vectors (see, for instance,
\cite{Hawk96,Boot99,Brow02})
\begin{equation}\label{ntilda}
\tilde n^i=\lambda \left( n^i-\eta u^i \right) ,\quad \tilde
u^i=\lambda\left( u^i+\eta n^i \right) \ ,
\end{equation}
where $n^i$ is understood as the normal to $\partial M_s$, the
variable $\eta =u^i n_i$ measures the non-orthogonality and
$\lambda =1/\sqrt{1+\eta ^2}$. For $\eta =0$ the foliation
surfaces are orthogonal to the boundary $\partial M_s$. Note that
the quantity $\lambda $ is connected to the proper radial velocity
$v$ by the relation \cite{Brow02} $\lambda =\sqrt{1-v^2}$.

In the present paper we deal with a real scalar field $\varphi
(x)$ propagating on the manifold $M$. The action describing the
dynamics of the field we will take in the form
\begin{equation}\label{action}
  S=S_{b}+S_{s},
\end{equation}
where the first term on the right,
\begin{equation}\label{actionB}
  S_b=\frac{1}{2}\int _{M}d^{D+1}x\sqrt{|g|}\left( \nabla _i \varphi \nabla ^i
  \varphi -m^2 \varphi ^2- \zeta R \varphi ^2\right) ,
\end{equation}
is the standard bulk action for the scalar field with curvature
coupling parameter $\zeta $, $R$ is the Ricci scalar for the
manifold $M$, $\nabla _i$ is the covariant derivative operator
associated with the bulk metric $g_{ik}$ (we adopt the conventions
of Ref. \cite{Birr82} for the metric signature and the curvature
tensor). The second term on the right of Eq. (\ref{action}),
\begin{equation}\label{actionGH}
  S_{s}=-\epsilon \int _{\partial M}d^Dx\sqrt{|h|}\left( \zeta \varphi
  ^2K +m_s \varphi ^2 \right),
\end{equation}
is the surface action with a parameter $m_s$. In Eq.
(\ref{actionGH}), $h$ is the determinant of the induced metric
$h_{ik}$ for the boundary $\partial M$,
\begin{equation}\label{indmetric}
  h_{ik}=g_{ik}-\epsilon n_in_k,
\end{equation}
which acts as a projection tensor onto $\partial M$, $K$ is the
trace of the extrinsic curvature tensor $K_{ik}$ of the boundary
$\partial M$:
\begin{equation}\label{extcurv}
  K=g^{ik}K_{ik}=h^{ik}K_{ik},\quad K_{ik}=h^l_ih^m_k\nabla _ln_m,\quad
  h^l_i=g^{lk}h_{ik}.
\end{equation}
For the most important special cases of minimally and conformally
coupled scalars in Eq. (\ref{actionB}) we have $\zeta =0,
(D-1)/4D$, respectively. Note that in Eq. (\ref{actionGH}) the
term with the extrinsic curvature scalar corresponds to the
Gibbons-Hawking surface term in general relativity, $-\epsilon
\int _{\partial M}d^Dx\sqrt{|h|}\, K$. This term is required so
that the action yields the correct equations of motion subject
only to the condition that the induced metric on the boundary is
held fixed. Note that for non-smooth boundary hypersurfaces it is
necessary to add to action (\ref{action}) the corner terms (for
these terms in the gravitational action of general relativity see
\cite{Hayw93,Hawk96b,Boot99,Brow02}) with integrations over
$(D-1)$-hypersurfaces at which the unit normal $n^i$ changes
direction discontinously. As we have assumed a smooth boundary
$\partial M_s$, in our case these hypersurfaces are presented by
the $(D-1)$-surfaces $\partial \Sigma _1=\Sigma _1\cap \partial
M_s$ and $\partial \Sigma _2=\Sigma _2\cap
\partial M_s$ and the corresponding corner term in the action
has the form
\begin{equation}\label{corneraction}
S_c=-\zeta \int_{\partial \Sigma _1}^{\partial \Sigma
_2}d^{D-1}x\sqrt{|\sigma |}\, \varphi ^2 \, {\mathrm{arcsinh}}
(\eta ),
\end{equation}
where $\int_{\partial \Sigma _1}^{\partial \Sigma
_2}=\int_{\partial \Sigma _2}-\int_{\partial \Sigma _1}$ and
$\sigma $ is the determinant of the metric tensor
\begin{equation}\label{sigmaik}
\sigma _{ik}=g_{ik} + n_i n_k - \tilde u_i \tilde u_k = g_{ik} +
\tilde n_i \tilde n_k -  u_i u_k
\end{equation}
induced on $\partial \Sigma $. In this paper we are interested in
the variation of the action with respect to the variations for
which the metric and field are held fixed on the initial and final
spacelike hypersurfaces $\Sigma _1$ and $\Sigma _2$, and
consequently on $\partial \Sigma _1$ and $\partial \Sigma _2$. In
this case the corner terms will not contribute to the variations
under consideration and we will omit them in the following
consideration (see, however, the discussion after formula
(\ref{varS2}) in Sec. \ref{sec:emtbs} for the total divergence
term in the variation of the action, located on the boundary).
Note that the contribution of the corner terms will be important
in the case of a non-smooth boundary $\partial M_s$ having
wedge-shaped parts.

\section{Variation of the action and the energy-momentum tensor}\label{sec:emtbs}

\subsection{Variation of the bulk action and the volume energy-momentum
tensor}\label{subsec:bactionvar}

In this subsection we consider the variation of the bulk action
(\ref{actionB}) induced by an infinitesimal variation $\delta
g_{ik}$ in the metric tensor. As the careful analysis of the
surface terms is crucial for our discussion we will present the
calculations in detail (for the corresponding calculations in
general relativity see, for instance, \cite{Brow02}). To find the
expression for the energy-momentum tensor of the scalar field with
action (\ref{action}), we need the variation of this action
induced by an infinitesimal variation $\delta g_{ik}$ in the
metric tensor.

First of all we consider the variation of the term involving the
Ricci scalar:
\begin{equation}
\delta _{g}\int_{M}d^{D+1}x\sqrt{|g|}\, \varphi ^{2}R=\int_{M}d^{D+1}x\sqrt{|g|}%
\, \varphi ^{2}\left( \delta g^{ik}G_{ik}+g^{ik}\delta
R_{ik}\right) , \label{varR1}
\end{equation}
where $R_{ik}$ and $G_{ik}=R_{ik}-g_{ik}R/2$ are Ricci and
Einstein tensors, respectively. To evaluate the second term in the
braces on the right we note that the contracted variation
$g^{ik}\delta R_{ik}$ is a pure spacetime divergence,
\begin{equation}
g^{ik}\delta R_{ik}=-\nabla _{l}W^{l},\quad W^{l}=g^{ik}\delta
\Gamma _{ik}^{l}-g^{il}\delta \Gamma _{ik}^{k},  \label{Wl}
\end{equation}
with $\Gamma _{ik}^{l}$ being the Christoffel symbol. In order to
find the variations of the Christoffel symbol in this formula,
note that from the identity $\delta (\nabla _{i}g_{kl})=0$ it
follows
\begin{equation}
g^{ik}\delta \Gamma _{ik}^{l}=-g^{il}\delta \Gamma
_{ik}^{k}+g^{lk}g^{ip}\nabla _{p}\delta g_{ik}.  \label{relvarW1}
\end{equation}
Combining this with the relations
\begin{equation}
\delta \Gamma _{ik}^{k}=\nabla _{i}(\delta \ln \sqrt{|g|})=-\frac{1}{2}%
g_{lm}\nabla _{i}(\delta g^{lm}),\quad g^{lk}g^{ip}\nabla
_{p}\delta g_{ik}=-\nabla _{m}(\delta g^{lm}),  \label{relvarW3}
\end{equation}
one finds
\begin{equation}
W^{l}=g^{il}g_{km}\nabla _{i}(\delta g^{km})-\nabla _{m}(\delta
g^{lm}). \label{W2}
\end{equation}

By taking into account formula (\ref{Wl}) and using the Stoke's
theorem
\begin{equation}
\int_{M}d^{D+1}x\sqrt{|g|}\, \nabla _{i}V^{i}=-\epsilon \int_{\partial M}d^{D}x%
\sqrt{|h|}\, n_{i}V^{i},  \label{Stoke}
\end{equation}
for the term with the variation of the Ricci tensor
one has
\begin{equation}
\int_{M}d^{D+1}x\sqrt{|g|}\, \varphi ^{2}g^{ik}\delta R_{ik}=\int_{M}d^{D+1}x%
\sqrt{|g|}\, W^{l}\nabla _{l}\varphi ^{2}+\epsilon \int_{\partial M}d^{D}x\sqrt{%
|h|}\, \varphi ^{2}n_{l}W^{l}.  \label{varRik1}
\end{equation}
For the first term on the right, using formula (\ref{W2}), we find
\begin{eqnarray}
\int_{M}d^{D+1}x\sqrt{|g|}\, W^{l}\nabla _{l}\varphi ^{2} &=&\int_{M}d^{D+1}x%
\sqrt{|g|}\, \delta g^{ik}\left( \nabla _{i}\nabla _{k}\varphi
^{2}-g^{lm}g_{ik}\nabla _{l}\nabla _{m}\varphi ^{2}\right)
\nonumber  \\
&&+\epsilon \int_{\partial M}d^{D}x\sqrt{|h|}\, \delta
g^{ik}\left( n_{i}\nabla _{k}\varphi ^{2}-g_{ik}n^{l}\nabla
_{m}\varphi ^{2}\right) .  \label{intWvar1}
\end{eqnarray}
Substituting this into formula (\ref{varRik1}) and taking into
account formula (\ref{W2}), for the term containing the variation
of the Ricci scalar, Eq. (\ref{varR1}), we obtain the formula
\begin{eqnarray}
\delta _{g}\int_{M}d^{D+1}x\sqrt{|g|}\, \varphi ^{2}R &=&\int_{M}d^{D+1}x\sqrt{%
|g|}\, \delta g^{ik}\left( \varphi ^{2}G_{ik}-g_{ik}g^{lm}\nabla
_{l}\nabla _{m}\varphi ^{2}+\nabla _{i}\nabla _{k}\varphi
^{2}\right)  \nonumber
 \\
&&+\epsilon \int_{\partial M}d^{D}x\sqrt{|h|}\left\{ \delta
g^{ik}\left( n_{i}\nabla _{k}\varphi ^{2}-g_{ik}n^{l}\nabla
_{l}\varphi ^{2}\right)
\right.    \label{varRiccisc} \\
&&+\left. \varphi ^{2}\left[ g_{km}n^{i}\nabla _{i}(\delta
g^{km})-n_{l}\nabla _{m}(\delta g^{lm})\right] \right\} .
\nonumber
\end{eqnarray}
This leads to the following variation of the bulk action
\begin{eqnarray}
\delta _{g}S_{b} &=&\frac{1}{2}\int_{M}d^{D+1}x\sqrt{|g|}\, \delta
g^{ik}T_{ik}^{\mathrm{(vol)}}+\frac{\epsilon }{2}\int_{\partial M}d^{D}x\sqrt{|h|}%
\left\{ \delta g^{ik}\left( -\zeta n_{i}\nabla
_{k}\varphi ^{2} \right. \right.   \nonumber  \\
&&+\left. \left. \zeta g_{ik}n^{l}\nabla _{l}\varphi ^{2}\right)
+\zeta \varphi ^{2}\left[ n_{l}\nabla _{m}(\delta
g^{lm})-g_{km}n^{i}\nabla _{i}(\delta g^{km})\right] \right\} ,
\label{varSB}
\end{eqnarray}
where $T_{ik}^{(\mathrm{vol})}$ is the standard metric
energy-momentum tensor for a scalar field with a curvature
coupling parameter $\zeta $ (see, for instance, \cite{Birr82}):
\begin{equation}
T_{ik}^{\mathrm{(vol)}}=\nabla _{i}\varphi \nabla _{k}\varphi -\frac{1}{2}%
g_{ik}g^{lm}\nabla _{l}\varphi \nabla _{m}\varphi
+\frac{m^2}{2}\varphi ^2 g_{ik}-\zeta \varphi ^{2}G_{ik}+\zeta
g_{ik}g^{lm}\nabla _{l}\nabla _{m}\varphi ^{2}-\zeta \nabla
_{i}\nabla _{k}\varphi ^{2}.  \label{Tik1}
\end{equation}
Note that this expression can be also written in the form
\begin{equation}
T_{ik}^{\mathrm{(vol)}}=T_{ik}^{(1)}+\frac{1}{2}\varphi
g_{ik}\left( g^{lm}\nabla _{l}\nabla _{m}\varphi +\zeta R\varphi
+m^2 \varphi \right) , \label{Tik11}
\end{equation}
where
\begin{equation}
T_{ik}^{(1)}=\nabla _{i}\varphi \nabla _{k}\varphi +\left[ \left( \zeta -%
\frac{1}{4}\right) g_{ik}g^{lm}\nabla _{l}\nabla _{m}-\zeta
R_{ik}-\zeta \nabla _{i}\nabla _{k}\right] \varphi ^{2} .
\label{Tik2}
\end{equation}
This tensor does not explicitly depend on the corresponding mass.
For the covariant divergences of tensors (\ref{Tik1}) and
(\ref{Tik2}) it can be seen that
\begin{eqnarray}\label{covdivvol}
 \nabla _i T_{k}^{{\mathrm{(vol)}}i}&=& \left( g^{lm}\nabla _l \nabla _m \varphi +\zeta R\varphi +
  m^2 \varphi \right) \nabla _k\varphi , \\
\nabla _i T_{k}^{(1)i}&=&-\frac{\varphi ^2}{2}\nabla _k\left[
\frac{1}{\varphi }\left( g^{lm}\nabla _l \nabla _m \varphi +\zeta
R\varphi +
  m^2 \varphi \right) \right] .\label{covdiv1}
\end{eqnarray}
From formula (\ref{varSB}) it follows that the surface term in the
variation of the bulk action contains the derivatives of the
variation for the metric. In next subsection we will see that
these terms are cancelled by the terms coming from the variation
of surface action (\ref{actionGH}).

\subsection{Variation of the surface action and the surface energy-momentum
tensor}\label{subsec:sactionvar}

Now we turn to the evaluation of the variation for surface action
(\ref {actionGH}) induced by the variation of the metric tensor.
This variation can be written in the form
\begin{equation}
\delta _{g}S_{s}=-\epsilon \int_{\partial M}d^{D}x\sqrt{|h|}\left[
\zeta \varphi ^{2}\delta _{g}K+\frac{1}{2}h^{ik}\delta g_{ik}
\left( \zeta \varphi ^2 K +m_s \varphi ^2 \right) \right] .
\label{varSGH1}
\end{equation}
First of all we note that for the variation of the extrinsic
curvature tensor trace one has
\begin{equation}
\delta _{g}K=\delta _{g}(g^{ik}\nabla _{i}n_{k})=\delta
g^{ik}\nabla _{i}n_{k}+g^{ik}\nabla _{i}\delta
n_{k}-n_{l}g^{ik}\delta \Gamma _{ik}^{l}. \label{varK1}
\end{equation}
By making use formulae
\begin{equation}
\delta n_{i}=\frac{1}{2}\epsilon n_{i}n^{k}n^{l}\delta
g_{kl},\quad \nabla _{i}n_{k}=K_{ik}+\epsilon n_{i}n^{l}\nabla
_{l}n_{k} , \label{varni}
\end{equation}
and formula (\ref{relvarW1}) for the contracted variation of the
Christoffel symbol, it can be seen that
\begin{equation}
\delta _{g}K=-K^{ik}\delta g_{ik}+\frac{\epsilon
}{2}Kn^{i}n^{k}\delta g_{ik}+n^{i}h^{kl}\left( \frac{1}{2}\nabla
_{i}\delta g_{kl}-\nabla _{k}\delta g_{il}\right) .  \label{varK2}
\end{equation}
Substituting this relation into formula (\ref{varSGH1}), for the
variation of the surface action one finds
\begin{eqnarray}
\delta _{g}S_{s}& =& -\epsilon \zeta \int_{\partial
M}d^{D}x\sqrt{|h|}\, \varphi
^{2}\left[ \delta g_{ik}\left( -K^{ik}+\frac{1}{2}Kh^{ik}+\frac{\epsilon }{2}%
Kn^{i}n^{k} + \frac{m_s}{2\zeta } h^{ik}\right) \right. \nonumber \\
&& \left. +n^{i}h^{kl}\left( \frac{1}{2}\nabla _{i}\delta
g_{kl}-\nabla _{k}\delta g_{il}\right) \right] . \label{varSGH2}
\end{eqnarray}
By taking into account that in Eq. (\ref{varSB}) we can replace
\begin{equation}
n_{l}\nabla _{m}(\delta g^{lm})-g_{km}n^{i}\nabla _{i}(\delta
g^{km})=g^{km}n^{l}\left[ \nabla _{l}(\delta g_{km})-\nabla
_{m}(\delta g_{lk})\right] ,  \label{rel1}
\end{equation}
and using formulae (\ref{varSB}) and (\ref{varSGH2}), for the
variation of the total action one finds
\begin{eqnarray}
\delta _{g}S &=&\frac{1}{2}\int_{M}d^{D+1}x\sqrt{|g|}\, \delta
g^{ik}T_{ik}^{\mathrm{(vol)}}+\frac{\epsilon }{2}\int_{\partial M}d^{D}x\sqrt{|h|}%
\left\{ \delta g^{ik}\left[ -\zeta n_{i}\nabla _{k}\varphi
^{2}+h_{ik}m_s \varphi ^2
\right. \right.   \nonumber \\
&&+\left. \left. \zeta g_{ik}n^{l}\nabla _{l}\varphi ^{2}-\zeta
\varphi ^{2}\left( 2K_{ik}-Kh_{ik}-\epsilon n_{i}n_{k}K\right)
\right] +\zeta \varphi ^{2}n^{i}h^{lm}\nabla _{l}(\delta
g_{im})\right\} .  \label{varS1}
\end{eqnarray}
Let us consider the last term in the figure braces on the right
which is the only term containing the derivatives of the metric
tensor variation on the boundary. To transform this term, we write
it in the form
\begin{equation}
n^{i}h^{lm}\nabla _{l}(\delta g_{im})=\nabla
_{l}(n^{i}h^{lm}\delta g_{im})-\delta g_{im}\nabla
_{l}(n^{i}h^{lm}).  \label{rel2}
\end{equation}
To proceed further we note that for an arbitrary vector field
$V^{l}$ one has
\begin{equation}
\nabla _{l}V^{l}=D_{l}V^{l}+\epsilon n^{i}n_{l}\nabla _{i}V^{l}=D
_{l}V^{l}+\epsilon n^{i}\nabla _{i}(n_{l}V^{l})-\epsilon
n^{i}V^{l}\nabla _{i}n_{l}, \label{rel3}
\end{equation}
with $D_{l}$ being the covariant derivative projected onto
$\partial M $. This allows us to write the first term on the right
of Eq. (\ref{rel2}) in the form
\begin{equation}
\nabla _{l}(n^{i}h^{lm}\delta g_{im})=D_{l}(n^{i}h^{lm}\delta
g_{im})-\epsilon n^{i}n^{k}h^{lm}\delta g_{km}\nabla _{i}n_{l},
\label{rel4}
\end{equation}
where we have taken into account that $n_{l}h^{lm}=0$. Now using
the second formula in Eq. (\ref{varni}), one finds
\begin{equation}
n^{i}h^{lm}\nabla _{l}(\delta g_{im})=D _{l}(n^{i}h^{lm}\delta
g_{im})-\delta g_{ik}\left( K^{ik}-\epsilon n^{i}n^{k}K\right) .
\label{rel5}
\end{equation}
Substituting this into formula (\ref{varS1}), we receive to the
final expression for the variation of the total action
\begin{eqnarray}
\delta _{g}S &=&\frac{1}{2}\int_{M}d^{D+1}x\sqrt{|g|}\, \delta
g^{ik}T_{ik}^{\mathrm{(vol)}}+\frac{\epsilon }{2}\int_{\partial
M}d^{D}x\sqrt{|h|} \left\{ \delta g^{ik}\left[ h_{ik}m_s \varphi
^2 \right. \right.   \nonumber
 \\
&&+\left. \left. \zeta h_{ik} n^{l}\nabla _{l}\varphi ^{2}-\zeta
\varphi ^{2}\left( K_{ik}-Kh_{ik}\right) \right] +\zeta
D_{l}(\varphi ^{2}n^{i}h^{lm}\delta g_{im})\right\} .
\label{varS2}
\end{eqnarray}
An important feature of this form is that the derivatives of the
metric tensor variation are contained in the form of a total
divergence on the boundary (the last total divergence term in the
figure braces). To see the relation between this total divergence
term and the variation of the corner term (\ref{corneraction}) in
the action, note that using formula (\ref{varni}) for the
variation $\delta n_i$, it can be seen that $h^{lm}n^i \delta
g_{im}=-h^l_k \delta n^k$. With this replacement, decomposing the
integral with the total divergence term into integrals over
$\partial M_s$, $\Sigma _1$, $\Sigma _2$ and using the Stoke's
theorem, we can see that
\begin{eqnarray}\label{reltocorneract}
\frac{\epsilon }{2} \int_{\partial M}d^Dx \sqrt{|h|}\, D_l(\varphi
^2 n^i h^{lm} \delta g_{im}) &=& \frac{1}{2}\int _{\partial \Sigma
_1}^{\partial \Sigma _2}d^{D-1}x \sqrt{|\sigma |}\, \varphi ^2
(\tilde u_k\delta n^k
+\tilde n_k\delta u^k) \nonumber \\
&=& \int _{\partial \Sigma _1}^{\partial \Sigma _2}d^{D-1}x
\sqrt{|\sigma |}\, \varphi ^2 \delta \, {\mathrm{arcsinh}}(\eta ).
\end{eqnarray}
In obtaining the second equality we have used relations
(\ref{ntilda}) (for similar calculations in general relativity
see, for instance, \cite{Brow02}). Now we see that the total
divergence term is cancelled by the corresponding term coming from
the variation of the corner part of the action, Eq.
(\ref{corneraction}).

Now we return to the variation given by Eq. (\ref{varS2}). Using
the Stoke's theorem, we can transform the integral with the total
divergence term into the integrals over $(D-1)$-surfaces $\partial
\Sigma _1$ and $\partial \Sigma _2$. We consider variations for
which the metric is held fixed on the initial and final spacelike
hypersurfaces $\Sigma _1$ and $\Sigma _2$ and, hence, on $\partial
\Sigma _1$ and $\partial \Sigma _2$. Consequently, for the class
of metric variations $\delta g^{ik}$ with $\delta g^{ik}|_{\Sigma
_j}=0$, $j=1,2$, the variation of the action takes the form
\begin{equation}\label{varS3}
\delta _{g}S =\frac{1}{2}\int_{M}d^{D+1}x\sqrt{|g|}\, \delta
g^{ik}T_{ik}^{\mathrm{(vol)}}+ \frac{1}{2}\int_{\partial
M_s}d^{D}x\sqrt{|h|}\, \delta g^{ik}\tau _{ik} \ ,
\end{equation}
where we have introduced a notation
\begin{equation}\label{tauik}
\tau _{ik}= \zeta \varphi ^{2} K_{ik} -h_{ik}\left( \zeta \varphi
^{2}K + \zeta n^{l}\nabla _{l}\varphi ^{2}+ m_s \varphi ^2\right)
\ ,
\end{equation}
and have taken into account that $\epsilon =-1$ for the timelike
hypersurface $\partial M_s$.

Using the definition of the metric energy-momentum tensor as the
functional derivative of the action with respect to the metric,
one finds
\begin{equation}
T_{ik}=\frac{2}{\sqrt{|g|}}\frac{\delta S}{\delta g^{ik}}%
=T_{ik}^{\mathrm{(vol)}}+T_{ik}^{\mathrm{(surf)}}. \label{Tikm1}
\end{equation}
Here the part of the energy-momentum tensor located on the surface
is defined by relation
\begin{equation}\label{surfemtnew}
T_{ik}^{\mathrm{(surf)}}= \delta (x;\partial M_s) \tau _{ik} \ ,
\end{equation}
and the 'one-sided' $\delta $-function $\delta (x;\partial M_s)$
is defined as
\begin{equation}
\int_{M}d^{D+1}x\sqrt{|g|}\, \delta (x;\partial M_s)=\int_{\partial M_s}d^{D}x\sqrt{%
|h|}.  \label{dfunc1}
\end{equation}
An important property of the surface energy-momentum tensor is
that it is orthogonal to the normal vector field $n^i$:
\begin{equation}\label{surforth}
n^i T_{ik}^{\mathrm{(surf)}}=0 \ ,
\end{equation}
which directly follows from the expression for the tensor $\tau
_{ik}$. Another important point is that, the surface
energy-momentum tensor does not depend on the foliation of
spacetime $M$ into spacelike hypersurfaces $\Sigma $. Using
expressions (\ref{Tik1}) and (\ref{surfemtnew}) for the volume and
surface energy-momentum tensors and the Gauss-Codacci equation
\begin{equation}\label{GaussCod}
  h^{l}_{m}\nabla _{l}K=h^{i}_{m}h^{kl}\nabla
  _{l}K_{ik}+h^{i}_mn^kR_{ik},
\end{equation}
the following relation can be derived
\begin{equation}
D_k \tau _m^k-n_kh^i_mT_i^{{\mathrm{(vol)}}k}=h^i_m\left( h^l_k
\nabla _l\tau _i^k-n_k T_i^{{\mathrm{(vol)}}k}\right) =
-h^i_m\left( 2\zeta \varphi K+  2m_s \varphi +n^l\nabla _l\varphi
\right) \nabla _i\varphi \label{covdivsurf}
\end{equation}
on the boundary $\partial M_s$.

Note that in all relations obtained so far we have not used the
field equation for $\varphi (x)$. Now we turn to this equation and
corresponding boundary condition  for the field following from the
extremality of the action with respect to variations in the scalar
field. The variation of action (\ref{action}) induced by an
infinitesimal variation $\delta \varphi $ in the field has the
form
\begin{eqnarray}
\delta _{\varphi }S&=&-\int_{M}d^{D+1}x\sqrt{|g|}\, \delta \varphi
\left[ g^{lm}\nabla _{l}\nabla _{m}\varphi +m^2 \varphi +\zeta
R\varphi \right]  \nonumber \\
&& -\epsilon \int _{\partial M}d^{D}x\sqrt{|h|}\, \delta \varphi
\left( 2\zeta \varphi K +n^l\nabla _l \varphi + 2m_s \varphi
\right) . \label{varphiS}
\end{eqnarray}
For the variations with $\delta \varphi =0$ on the initial and
final hypersurfaces $\Sigma _1$ and $\Sigma _2$, this leads to the
following field equation on the bulk
\begin{equation}
g^{lm}\nabla _{l}\nabla _{m}\varphi +m^2 \varphi +\zeta R\varphi
=0, \label{fieldeqphi}
\end{equation}
and to the boundary condition of Robin type,
\begin{equation}\label{boundcondquadVs}
2(\zeta K +m_s)\varphi + n^l\nabla _l \varphi =0 ,
\end{equation}
on the hypersurface $\partial M_s$. Note that the Dirichlet
boundary condition is obtained from here in the limit $m_s\to
\infty $.

From relation (\ref{Tik11}) for the tensors
$T_{ik}^{{\mathrm{(vol)}}}$ and $T^{(1)}_{ik}$ we see that on the
class of solutions to field equation (\ref{fieldeqphi}) these
tensors coincide: $T_{ik}^{\mathrm{(vol)}} = T_{ik}^{(1)}$. An
advantage of the form $T^{(1)}_{ik}$ in calculations of the vacuum
expectation values of the energy density in the corresponding
quantum theory is that it is sufficient to evaluate only two
quantities, $\langle 0|\varphi ^2(x)|0\rangle$ and $\langle
0|(\partial _0\varphi )^2|0\rangle$. In the case of
$T_{ik}^{{\mathrm{(vol)}}}$, due to the second term on the right
of Eq. (\ref{Tik1}), we need the vacuum expectation values for all
derivatives $\langle 0|\partial _l \varphi
\partial _m \varphi |0\rangle$. In fact, this is the reason for our
choice of the form $T^{(1)}_{ik}$ in investigations of the Casimir
densities for various types of geometries. Note that, using
boundary condition (\ref{boundcondquadVs}), the surface
energy-momentum tensor can be also written in the form
\begin{equation}
T_{ik}^{\mathrm{(surf)}}=  \delta (x;\partial M_s)\left[ \zeta
\varphi ^2K_{ik}- \left( 2\zeta -\frac{1}{2} \right) h_{ik}
\varphi n^l\nabla _l\varphi \right] . \label{surfemtform2}
\end{equation}
The surface energy-momentum tensor in this form does not
explicitly depend on the parameter $m_s$. From relations
(\ref{covdivvol}) and (\ref{covdivsurf}), by making use field
equation (\ref{fieldeqphi}) and boundary condition
(\ref{boundcondquadVs}), we see that on the solutions of the field
equation the volume energy-momentum tensor covariantly conserves,
\begin{equation}\label{covdiv0}
 \nabla _i T_{k}^{{\mathrm{(vol)}}i}=0,
\end{equation}
and for the surface energy-momentum tensor one has
\begin{equation}\label{covdivsurf0}
D_i \tau _{k}^{i}=n_lh^i_kT_{i}^{{\mathrm{(vol)}}l} \ .
\end{equation}
The last equation expresses the local conservation of the boundary
energy-momentum tensor up to the flow of the energy-momentum
across the boundary into $M$. Integral consequences from formulae
(\ref{covdiv0}) and (\ref{covdivsurf0}) we will consider in Sec.
\ref{sec:intcons}.

The traces for the volume and surface energy-momentum tensors are
easily obtained from (\ref{Tik2}) and (\ref{surfemtform2}):
\begin{eqnarray}
T_{i}^{{\mathrm{(vol)}}i} &=&D(\zeta -\zeta _{c})g^{lm}\nabla
_l\nabla _m \varphi ^2+m^2 \varphi ^2,
\label{Tiktraces1} \\
T_{i}^{{\mathrm{(surf)}}i} &=& - \delta (x;\partial M_s)\left[
D(\zeta -\zeta _{c})n^l\nabla _l\varphi ^2+m_s \varphi ^2 \right],
\label{Tiktraces2}
\end{eqnarray}
where $\zeta _{c}=(D-1)/4D$ is the curvature coupling parameter
for a conformally coupled scalar field. Hence, for a conformally
coupled massless scalar with $m_s =0$ both these traces vanish.
The latter property also directly follows from the conformal
properties of the action (\ref{action}). Indeed, it can be seen
that under conformal transformations of the metric $\bar
g_{ik}=\Omega ^2(x) g_{ik}$ and the corresponding transformation
of the field $\bar \varphi =\Omega ^{(1-D)/2}\varphi $, action
(\ref{action}) with $m=m_s=0$, transforms as
\begin{equation}\label{Sconftrans}
S[\bar g_{ik},\bar \varphi ] = S[g_{ik}, \varphi ] +D(\zeta -\zeta
_c)\int_{M}d^{D+1}x\sqrt{|g|}\left[ 2\varphi \nabla ^{i}\varphi
\nabla _{i} \ln \Omega -\frac{D-1}{2}\varphi ^2 \nabla _i \ln
\Omega \nabla ^{i}\ln \Omega \right] ,
\end{equation}
and, hence, is conformally invariant in the case $\zeta =\zeta
_c$. In order to obtain (\ref{Sconftrans}) we have used the
standard transformation formula for the Ricci scalar and the
corresponding formula for the extrinsic curvature tensor:
\begin{equation}\label{Kikconftrans}
\bar K_{ik}=\Omega K_{ik} + h_{ik}n^l \nabla _l \Omega \ .
\end{equation}
From the conformal invariance of the action it follows that
boundary condition (\ref{boundcondquadVs}) is also conformally
invariant when $m_s=0$ and $\zeta =\zeta _c$. In this case it
coincides with the conformally invariant Hawking boundary
condition \cite{Kenn80}. This is also directly seen from the
relation
\begin{equation}\label{boundconftrans}
2\zeta \bar K \bar \varphi +\bar n^l \nabla _l \bar \varphi =
\Omega ^{-\frac{D+1}{2}}\left[ 2\zeta K \varphi +n^l \nabla _l
\varphi +2D(\zeta -\zeta _c)\varphi n^l \nabla _l \ln \Omega
\right] \ ,
\end{equation}
where $\bar n^l=\Omega ^{-1}n^l$. Note that for general $m_s$,
under the conformal transformations the surface energy-momentum
tensor transforms as
\begin{eqnarray}\label{Tsurfconftrans}
\bar T_i^{{\mathrm{(surf)}}k}[\bar \varphi (x)] &=& \bar \delta
(x;\partial M_s) \bar \tau _i^k [\bar \varphi (x)] \nonumber \\
&=& \Omega ^{-D-1}\left\{ T_i^{{\mathrm{(surf)}}k}[\varphi (x)] -D
(\zeta -\zeta _c) \delta (x;\partial M_s) h_i^k \varphi ^2 n^l
\nabla _l \ln \Omega \right\} ,
\end{eqnarray}
where $\bar \delta (x;\partial M_s) =\Omega ^{-1} \delta
(x;\partial M_s)$. For $\zeta =\zeta _c$ this transformation
formula is the same as for the volume part of a conformally
coupled massless scalar field.

As an example let us consider the special case of a static
manifold $M$ with static boundary $\partial M_s$. In an
appropriately chosen coordinate frame one has $g_{0\beta }=0$ and
the components $g_{00}$, $g_{\beta \rho }$, $\beta , \rho
=1,2,\ldots , D$, do not depend on the time coordinate. Taking
into account that in static coordinates $n^0=0$, for the
$00$-component of the extrinsic curvature tensor we find
$K_{00}=\frac{1}{2}n^{\beta }
\partial _{\beta }g_{00}$. Substituting this into Eq.
(\ref{surfemtform2}), for the surface energy density we obtain the
expression
\begin{eqnarray}\label{Tsurf00st}
T^{{\mathrm{(surf)}}0}_0 &=&\delta (x;\partial M_s) \left[ \zeta
\varphi ^2 n^{\beta }\partial _{\beta }(\ln \sqrt{g_{00}})- \left(
2\zeta -\frac{1}{2}\right) \varphi n^\beta
\partial _{\beta }\varphi \right] \nonumber \\
&=& \zeta \varphi ^2\delta (x;\partial M_s) \left[ 4\zeta n^{\beta
}\partial _{\beta }(\ln \sqrt{g_{00}}) +(4\zeta -1) (K_\beta
^\beta +m_s/\zeta )\right] ,
\end{eqnarray}
where in the second line we have used boundary condition
(\ref{boundcondquadVs}). In the particular case $g_{00}=1$ the
first expression coincides with the formula of the surface energy
density used in Ref. \cite{Kenn80} to cancel the divergence in the
vacuum expectation value of the local energy density for a
quantized scalar field and in Refs. \cite{Rome02,Saha01,Rome01} to
evaluate the vacuum expectation values for flat, spherical and
cylindrical static boundaries on the Minkowski background. Note
that in this case the surface energy vanishes for Dirichlet and
Neumann boundary conditions. A simple example when the term with
$g_{00}$ in Eq. (\ref{Tsurf00st}) gives contribution and the
formula used in the papers cited above is not applicable,
corresponds to an infinite plane boundary moving with uniform
proper acceleration $a^{-1}$ in the Minkowski spacetime. Assuming
that the boundary is situated in the right Rindler wedge, in the
accelerated frame it is convenient to introduce Rindler
coordinates $(\tau ,\xi ,{\bf x})$ related to the Minkowski ones
$(t ,x^1 ,{\bf x})$ by transformations
\begin{equation}\label{RindtoMink}
t=\xi \sinh \tau , \quad x^1= \xi \cosh \tau ,
\end{equation}
and ${\bf x} =(x^2,\ldots ,x^D)$ denotes the set of coordinates
parallel to the boundary. In these coordinates the metric is
static, admitting the Killing vector field $\partial /\partial
\tau $ and has the form
\begin{equation}\label{Rindmetric}
g_{ik}={\mathrm{diag}} (\xi ^2,-1,\ldots ,-1).
\end{equation}
The boundary $\partial M_s$ is defined by $\xi =a$. For the
boundary normal and nonzero components of the extrinsic curvature
tensor one has
\begin{equation}
n^{i}=\delta _{1}^{i},\quad K_{00}=\xi ,\quad K=1/\xi \ .
\label{nK00Rind}
\end{equation}
Now for the components of the tensor $\tau _i^k$ in expression
(\ref{surfemtnew}) for the surface energy-momentum tensor we find
\begin{equation}\label{Rindtau1}
   \tau _i^k = \frac{\zeta }{a} \varphi ^2 \left[ \delta _i^0 \delta _0^k
   +\left( 4\zeta -1\right) (1+m_s a/\zeta )\delta _i^k \right]
   , \quad i,k=0,2\ldots , D,
\end{equation}
and $\tau _1^1=0$. As we see from here, now for non-minimal
coupling the surface energy density does not vanish for Neumann
boundary condition (realized for $m_s=-\zeta /a$). An application
of formula (\ref{Rindtau1}) will be given in Sec.
\ref{sec:vacemt}.

\section{Integral conservation laws} \label{sec:intcons}

Integral conservation laws are associated with the symmetries of
the background manifold. These symmetries are described by the
Killing vector $\eta _{i}$ satisfying the equation
\begin{equation}
\nabla _{i}\eta _{k}+\nabla _{k}\eta _{i}=0.  \label{Kileq}
\end{equation}
By virtue of the symmetry of the tensor $T_{ik}^{{\rm (vol)}}$,
from equation (\ref{covdiv0}) one can obtain
\begin{equation}
\frac{1}{\sqrt{|g|}}\partial _{i}\left( \sqrt{|g|}T^{{\rm
(vol)}ik}\eta _{k}\right) =0.  \label{divzero}
\end{equation}
Integrating this equation over the spacetime region $M$ and using
the Stoke's theorem one obtains the relation
\begin{equation}
P^{{\rm (vol)}}_{\Sigma _2} - P^{{\rm (vol)}}_{\Sigma _1} = -
\int_{\partial M_s} d^D x \sqrt{|h|} \, n_i \eta _k T^{{\rm
(vol)}ik}, \label{Pvolchange}
\end{equation}
where
\begin{equation}
P^{{\rm (vol)}}_{\Sigma }=\int_{{\Sigma}}d^{D}x\sqrt{|\gamma |} \,
u_i \eta _{k} T^{{\rm (vol)}ik} \label{Pvol}
\end{equation}
is the volume part of the energy-momentum, and $u^i$ is the future
pointing normal vector to the spatial hypersurface $\Sigma $. In
Eq. (\ref{Pvol}), $\gamma $ is the determinant for the metric
$\gamma _{ik}$ induced on the surfaces $\Sigma $ by the spacetime
metric: $\gamma _{ik}=g_{ik}-u_i u_k$. As we see from
(\ref{Pvolchange}), the volume energy-momentum, in general, is not
conserved separately.

Projecting the Killing vector onto the boundary we find the
corresponding vector for the boundary geometry. We will denote
this vector by the index $b$:
\begin{equation}\label{Kilbound}
\eta _i^{(b)}=h^{k}_{i}\eta _k.
\end{equation}
The equation satisfied by this vector field is obtained as the
projection of Eq. (\ref{Kileq}):
\begin{equation}\label{Kileqb}
D_i\eta ^{(b)}_{k}+D_k\eta ^{(b)}_{i}=0.
\end{equation}
In order to construct global quantities let us write equation
(\ref{covdivsurf0}) in the form
\begin{equation}\label{covdivsurf2}
D_k\tau ^{ik}=h^{i}_k n_l T^{{\mathrm{(vol)}}kl} \ .
\end{equation}
Multiplying this equation by $\eta ^{(b)}_i$, using the symmetry
properties of the surface energy-momentum tensor and Eq.
(\ref{Kileqb}), and noting that from the orthogonality relation
$n_i\tau ^{ik}=0$ one has $\eta ^{(b)}_i\tau ^{ik}=\eta _i\tau
^{ik}$, we find
\begin{equation}\label{sovdivsurf}
\partial _{k}\left( \sqrt{|h|}\eta _{i}\tau ^{ik}\right) =\sqrt{|h|}
h_k^i\eta _i^{(b)} n_l T^{{\mathrm{(vol)}}kl} \ .
\end{equation}
We integrate this relation over the  hypersurface $\partial M_s$.
After using the Stoke's theorem for the term on the left and
noting that $h_k^i\eta _i^{(b)}=\eta _k^{(b)}$, one receives
\begin{equation}\label{Psurfchange}
P^{{\rm (surf)}}_{\partial \Sigma _2} - P^{{\rm (surf)}}_{\partial
\Sigma _1} =  \int_{\partial M_s} d^D x \sqrt{|h|}\, n_i \eta
_k^{(b)} T^{{\rm (vol)}ik} \ .
\end{equation}
Here we have introduced the surface energy-momentum
\begin{equation}\label{Psurf}
P^{{\rm (surf)}}_{\partial \Sigma } = \int_{{\partial \Sigma}}
d^{D-1}x \sqrt{|\sigma |} \, \eta _{k}\tilde u_{i} \tau ^{ik},
\end{equation}
where $\tilde u_{i}$ is the forward pointing unit normal to the
boundary $\partial \Sigma $ in the hypersurface $\partial M_s$ and
has been introduced in Sec. \ref{sec:emtbs}, $\sigma $ is the
determinant of the metric (\ref{sigmaik}) induced on $\partial
\Sigma $.

Now summing the relations (\ref{Pvolchange}) and
(\ref{Psurfchange}), we arrive at the formula
\begin{equation}
P_{\Sigma _2} - P_{\Sigma _1} = \int_{\partial M_s} d^D x
\sqrt{|h|} \, n_i n_k n^l \eta _l T^{{\rm (vol)}ik} \ .
\label{Ptotchange}
\end{equation}
Here $P_{\Sigma }$ is the total energy-momentum on a hypersurface
$\Sigma $ being the sum of volume and surface parts:
\begin{equation}\label{Ptot}
P_{\Sigma } = P^{{\rm (vol)}}_{\Sigma } + P^{{\rm
(surf)}}_{\partial \Sigma }.
\end{equation}
As it follows from (\ref{Ptotchange}), for any Killing vector
tangent to the hypersurface $\partial M_s$, $n^l \eta _l=0$, the
total energy-momentum $P_{\Sigma }$ is a conserved quantity. The
subintegrand on the right of formula (\ref{Ptotchange}) can be
interpreted in terms of the normal force acting on the boundary.
This force is determined by the volume part of the energy-momentum
tensor. The total energy-momentum can be also obtained from the
total energy-momentum tensor (\ref{Tikm1}) by integrating over the
hypersurface $\Sigma $:
\begin{equation}
P_{\Sigma }=\int_{{\Sigma}}d^{D}x\sqrt{|\gamma |} \, u_i \eta _{k}
T^{ik} \ . \label{Ptot1}
\end{equation}
To see this, first note that using the relations (see, for
instance, \cite{Hawk96}) $\sqrt{|g|}=N\sqrt{|\gamma |}$,
$\sqrt{|\gamma |}=N\lambda \sqrt{|\sigma |}$ with the lapse
function $N$ and taking infinitesimaly close $\Sigma _1$ and
$\Sigma _2$, from Eq. (\ref{dfunc1}) we obtain
\begin{equation}\label{deltasigma}
\int _{\Sigma }d^D x\sqrt{|\gamma |}\, \delta (x;\partial M_s) =
\int_{\partial \Sigma }d^{D-1}x\sqrt{|\sigma |} \, \lambda \ .
\end{equation}
From the orthogonality relation $n_i\tau ^{ik}=0$ and the second
formula in (\ref{ntilda}) we have
\begin{equation}\label{lamu}
\lambda u_i\tau ^{ik}=\tilde u_i\tau ^{ik} \ .
\end{equation}
Hence, by taking into account (\ref{surfemtnew}) and (\ref{lamu}),
the integral with the surface energy-momentum tensor gives
\begin{equation}\label{Psurf4}
\int_{\Sigma }d^D x \sqrt{|\gamma |}\, u_i\eta _k
T^{{\mathrm{(surf)}}ik}=\int_{\partial \Sigma
}d^{D-1}x\sqrt{|\sigma |}\, \lambda u_i\eta _k \tau
^{ik}=P_{\partial \Sigma }^{{\mathrm{(surf)}}} \ ,
\end{equation}
with $P_{\partial \Sigma }^{{\mathrm{(surf)}}}$ defined as in Eq.
(\ref{Psurf}).

To see how the quantities  $P_{\Sigma }^{{\mathrm{(vol)}}}$ and
$P_{\partial \Sigma }^{{\mathrm{(surf)}}}$ are related to the
corresponding proper densities (for a similar discussion in
general relativity see \cite{Brow93}), let us define the proper
energy volume density $\varepsilon ^{{\mathrm{(vol)}}}$, proper
momentum volume density $p^{{\mathrm{(vol)}}i}$, and spatial
stress $\rho ^{{\mathrm{(vol)}}ik}$ on the hypersurface $\Sigma $
as
\begin{equation}\label{volpropdens}
\varepsilon ^{{\mathrm{(vol)}}}=u_k u_l T^{{\mathrm{(vol)}}kl},
\quad p^{{\mathrm{(vol)}}i}=\gamma ^i_k u_l T^{{\mathrm{(vol)}}kl}
, \quad \rho ^{{\mathrm{(vol)}}ik}=\gamma ^i_l \gamma ^k_m
T^{{\mathrm{(vol)}}lm} \ .
\end{equation}
The following relation can be easily checked:
\begin{equation}\label{relTeps}
u_l T^{{\mathrm{(vol)}}il} = \varepsilon ^{{\mathrm{(vol)}}}u^i+
p^{{\mathrm{(vol)}}i} \ ,
\end{equation}
and, hence, expression (\ref{Pvol}) for the volume energy-momentum
can be written in terms of proper quantities as
\begin{equation}\label{Pvoleps}
P^{{\mathrm{(vol)}}}_{\Sigma }= \int _{\Sigma }d^D x \sqrt{|\gamma
|} \, \eta _k \left( \varepsilon ^{{\mathrm{(vol)}}}u^k
+p^{{\mathrm{(vol)}}k}\right) \ .
\end{equation}
Similarly, we can define the proper energy surface density
$\varepsilon ^{{\mathrm{(surf)}}}$, proper momentum surface
density $p^{{\mathrm{(surf)}}i}$, and spatial stress $\rho
^{{\mathrm{(surf)}}ik}$ on the surface $\partial \Sigma $:
\begin{equation}\label{surfpropdens}
\varepsilon ^{{\mathrm{(surf)}}}=\tilde u_k \tilde u_l \tau ^{kl},
\quad p^{{\mathrm{(surf)}}i}=\sigma ^i_k \tilde u_l \tau ^{kl} ,
\quad \rho ^{{\mathrm{(surf)}}ik}=\sigma ^i_l \sigma ^k_m \tau
^{lm} \ ,
\end{equation}
with the relation
\begin{equation}\label{reltaueps}
\tilde u_l \tau ^{il} = \varepsilon ^{{\mathrm{(surf)}}}\tilde
u^i+ p^{{\mathrm{(surf)}}i} \ .
\end{equation}
The latter allows us to write expression (\ref{Psurf}) for the
surface energy-momentum in the form
\begin{equation}\label{Psurfeps}
P^{{\mathrm{(surf)}}}_{\partial \Sigma }= \int _{\partial \Sigma
}d^{D-1} x \sqrt{|\sigma |} \, \eta _k \left( \varepsilon
^{{\mathrm{(surf)}}}\tilde u^k +p^{{\mathrm{(surf)}}k}\right) \ .
\end{equation}

If a Killing vector field $\eta _i=\eta _i^{(0)}$ exists which is
timelike and has unit length, $\eta ^{(0)i} \eta ^{(0)}_i=1$, we
can take a foliation $\Sigma $ with $u^i =\eta ^{(0)i}$. In this
case $\eta ^{(0)}_k p^{{\mathrm{(vol)}}k}=0$, and from
(\ref{Pvoleps}) one finds
\begin{equation}\label{Pvoleps2}
E^{{\mathrm{(vol)}}}_{\Sigma }= \int _{\Sigma }d^D x \sqrt{|\gamma
|} \, \varepsilon ^{{\mathrm{(vol)}}} \ ,
\end{equation}
which is the total proper energy on the hypersurface $\Sigma $.
Similarly, noting that $u_k \sigma ^k_l=0$ and, hence, $u_k
p^{{\mathrm{(surf)}}k}=0$, from (\ref{Psurfeps}) we receive
\begin{equation}\label{Psurfeps2}
E^{{\mathrm{(surf)}}}_{\partial \Sigma }= \int _{\partial \Sigma
}d^{D-1} x \sqrt{|\sigma |} \frac{\varepsilon
^{{\mathrm{(surf)}}}}{\sqrt{1-v^2}} \ ,
\end{equation}
where we have introduced the proper radial velocity $v$ by using
the relation $\sqrt{1+\eta ^2}=1/\sqrt{1-v^2}$ (see Sec.
\ref{sec:action}). For a static boundary $\partial \Sigma $ one
has $n^l \eta ^{(0)}_l=0$ and $\eta =v=0$ and, as it follows from
(\ref{Ptotchange}), the total energy does not depend on the
hypersurface $\Sigma $ and is a conserved quantity. Note that, in
general, this is not the case for volume and surface parts
separately.

\section{Vacuum expectation values for the energy-momentum tensor
on manifolds with boundaries and the vacuum energy}
\label{sec:vacemt}

In this section we will consider a quantum scalar field $\varphi
(x)$ on a static manifold $M$ with boundary $\partial M_s$. In a
static coordinate frame, $g_{0\beta }=0$ and the components
$g_{00}$, $g_{\beta \rho }$, $\beta , \rho =1,\ldots , D$, do not
depend on the time coordinate and $n^0=0$ on the boundary
$\partial M_s$. The quantization procedure more or less parallels
that in Minkowski spacetime. Let $\{\varphi _{\alpha }(x),\varphi
_{\alpha }^{\ast }(x)\}$ be a complete set of positive and
negative frequency solutions to the field equation, obeying the
boundary condition (\ref{boundcondquadVs}) on the boundary
$\partial M_s$. Here $\alpha $ corresponds to a set of quantum
numbers, specifying the eigenfunctions. These eigenfunctions are
orthonormalized in accordance with \cite{Birr82}
\begin{equation}\label{orthnorm}
  (\varphi _{\alpha },\varphi _{\alpha '})=-i\int _{\Sigma}d^Dx
  \sqrt{|\gamma |}u^k\left[ \varphi _{\alpha }^{\ast }(x)\partial _k
  \varphi _{\alpha '}(x) -\varphi _{\alpha '}(x)\partial _k
  \varphi _{\alpha }^{\ast }(x)\right] =\delta _{\alpha \alpha '},
\end{equation}
where the integration goes over a spatial hypersurface ${\Sigma}$
of the manifold $M$ with the future pointing normal $u^k$. It can
be seen that scalar product (\ref{orthnorm}) does not depend on
the choice of the hypersurface $\Sigma $. Indeed, for the
corresponding difference for two hypersurfaces $\Sigma _1$ and
$\Sigma _2$, from the field equation, by using the Stoke's
theorem, one finds
\begin{equation}\label{orthnorm1}
  (\varphi _{\alpha },\varphi _{\alpha '})_{\Sigma _1}-
  (\varphi _{\alpha },\varphi _{\alpha '})_{\Sigma _2}=-i
  \int _{\partial M_s}d^Dx
  \sqrt{|h|}n^k\left[ \varphi _{\alpha }^{\ast }(x)\partial _k
  \varphi _{\alpha '}(x) -\varphi _{\alpha '}(x)\partial _k
  \varphi _{\alpha }^{\ast }(x)\right] .
\end{equation}
Making use boundary condition (\ref{boundcondquadVs}) for the
functions $\varphi _{\alpha }(x)$, we conclude that the integral
on the right vanishes. Due to the staticity of the problem the
time dependence of the eigenfunctions can be taken in the form
$e^{i\omega _{\alpha }t}$, with $\omega _{\alpha }$ being the
corresponding eigenfrequencies. By expanding the field operator
over the eigenfunctions $\varphi _{\alpha }(x) $, using the
standard commutation rules and the definition of the vacuum state,
for the vacuum expectation values of the energy-momentum tensor
one obtains
\begin{equation}
\langle 0|T_{\mu \nu }(x)|0\rangle =\sum_{\alpha }T_{\mu \nu }\{\varphi {%
_{\alpha }(x),\varphi _{\alpha }^{\ast }(x)\}},  \label{emtvev1}
\end{equation}
where $|0\rangle $ is the amplitude for the corresponding vacuum
state, and the bilinear form $T_{\mu \nu }\{{\varphi ,\psi \}}$ on
the right is determined by the classical energy-momentum tensor.
The mode-sum on the right of formula (\ref{emtvev1}) is divergent.
Various regularization procedures can be used to make it finite
(cutoff function, zeta function regularization, dimensional
regularization, point-splitting technique). In the formulae below,
it is implicitly assumed that some regularization procedure is
used and the corresponding relations take place for regularized
quantities.

To find the total vacuum energy on $M$, we need the vacuum
expectation values of the energy density $\langle
0|T_{0}^{0}|0\rangle $. For the points $x\notin \partial M_s$ the
latter can be evaluated by the formula
\begin{equation}
\langle 0|T_{0}^{{\mathrm{(vol)}}0}|0\rangle =\sum_{\alpha }\left[
\omega _{\alpha }^{2}+(\zeta -1/4)g_{00}g^{lm}\nabla _{l}\nabla
_{m}+\zeta \Gamma _{00}^{\beta }\nabla _{\beta }-\zeta
R_{00}\right] \varphi _{\alpha }(x)\varphi _{\alpha }^{\ast }(x).
\label{vacvolendens}
\end{equation}
In the static case we have
\begin{equation}
\Gamma _{00}^{\beta }=\frac{1}{2}\tilde \gamma ^{\beta \rho }\tilde{\nabla }%
_{\sigma }g_{00},\quad R_{00}=-\frac{1}{\sqrt{g_{00}}}\widetilde{\nabla }%
_{\beta }\widetilde{\nabla }^{\beta }\sqrt{g_{00}},\quad \beta
,\rho =1,\ldots ,D,  \label{GamR}
\end{equation}
where $\tilde{\nabla }_{\beta }$ is the covariant derivative
operator associated with the $D$-dimensional spatial metric
$\tilde \gamma _{\beta \rho }=-g_{\beta \rho }$. For the nonzero
components of the timelike Killing vector $\eta ^{(0)i}$ and the
normal vector $u^i$ one has $\eta ^{(0)0}=1$,
$u^0=1/\sqrt{g_{00}}$. Substituting these into formula
(\ref{Pvol}) and by taking into account relations (\ref{GamR}),
for the total volume energy one finds
\begin{eqnarray}
E^{{\mathrm{(vol)}}} &=&\int_{{\Sigma}}d^{D}x\sqrt{|g|}\langle
0|T_{0}^{{\mathrm{(vol)}}0}|0\rangle \nonumber \\
&=&\sum_{\alpha }\left\{ \frac{\omega _{\alpha }}{2}+\int_{\partial {\Sigma}%
}d^{D-1}x\sqrt{|g|}n^{\beta }\left[ (\zeta -1/4)\partial _{\beta
}-\zeta (\partial _{\beta }\ln \sqrt{g_{00}})\right] \varphi
_{\alpha }(x)\varphi _{\alpha }^{\ast }(x)\right\} , \label{Evol}
\end{eqnarray}
where $\partial {\Sigma}$ is the boundary of ${\Sigma}$. As we
see, the total volume energy, in general, does not coincide with
the vacuum energy evaluated as the sum of zero-point energies for
each elementary oscillator. The total surface energy can be found
integrating the corresponding energy density from Eq.
(\ref{Tsurf00st}):
\begin{eqnarray}
E^{{\mathrm{(surf)}}} &=&\int_{{\Sigma}}d^{D}x\sqrt{|g|}\langle
0|T_{0}^{{\mathrm{(surf)}}0}|0\rangle \nonumber \\
&=&-\sum_{\alpha }\int_{\partial
{\Sigma}}d^{D-1}x\sqrt{|g|}n^{\beta }\left[ (\zeta -1/4)\partial
_{\beta }-\zeta (\partial _{\beta }\ln \sqrt{g_{00}})\right]
\varphi _{\alpha }(x)\varphi _{\alpha }^{\ast }(x). \label{Esurf}
\end{eqnarray}
For the total vacuum energy one obtains
\begin{equation}
E=E^{{\mathrm{(vol)}}}+E^{{\mathrm{(surf)}}}=\frac{1}{2}\sum_{\alpha
}\omega _{\alpha }. \label{Etot}
\end{equation}
As we have seen, the surface energy contribution is crucial to
obtain the equality between the vacuum energy evaluated as the sum
of zero-point energies of each elementary oscillator, and energy
obtained by the integration of the corresponding energy density.

As an example let us consider vacuum expectation value of the surface
energy-momentum tensor for a massless scalar field induced by an
infinite plane boundary moving with uniform proper acceleration
$a^{-1}$ through the Fulling-Rindler vacuum. The vacuum
expectation value  of the volume part of the energy-momentum
tensor for this problem is investigated in Refs.
\cite{Cand77,Saha02} (see Ref. \cite{Avag02} for the case of two
parallel plates geometry). The corresponding total Casimir
energies for a single and two plates cases are recently evaluated
in Ref. \cite{Saha03b} by using the zeta function technique. In
the consideration below we will assume that the plate is situated
in the right Rindler wedge and will consider the region on the
right from the plate (RR region in terminology of Ref.
\cite{Saha02}) corresponding to $\xi \geq a $ in the Rindler
coordinates (see Sec. \ref{sec:emtbs}). The boundary condition
(\ref{boundcondquadVs}) now takes the form
\begin{equation}
\left( A+a\frac{\partial }{\partial \xi }\right) \varphi =0,\quad
A\equiv 2 (\zeta +m_{s}a) ,\quad  \xi =a .  \label{boundRind}
\end{equation}
For the region $\xi \geq a$ a complete set of solutions to the
field equation that are of positive frequency with respect to
$\partial /\partial \tau $ and bounded as $\xi \rightarrow \infty
$ is
\begin{equation}
\varphi _{\alpha }(x)=CK_{i\omega }(k \xi )e^{i{\bf kx}-i\omega
\tau },\quad \alpha =(\omega ,{\bf k}),  \label{sol2}
\end{equation}
where $k=|{\mathbf{k}}|$ and $K_{i\omega }(x)$ is the MacDonald
function with the imaginary order. From boundary condition
(\ref{boundRind}) we find that the possible values for $\omega $
have to be roots to the equation
\begin{equation}
AK_{i\omega }(k a)+k a K_{i\omega }^{\prime }(k a)=0,
\label{modeeq}
\end{equation}
where the prime denotes the differentiation with respect to the
argument.
This equation has infinite number of real zeros. We will denote them by $%
\omega =\omega _{n}=\omega _{n}(k)$, $n=1,2,...$. The coefficient
$C$ in Eq. (\ref{sol2}) is determined by the normalization
condition and is equal to
\begin{equation}
C^{2}=\frac{1}{(2\pi )^{D-1}}\frac{\bar{I}_{i\omega _{n}}(k a)}{\frac{%
\partial }{\partial \omega }\bar{K}_{i\omega }(k a)\mid _{\omega
=\omega _{n}}},  \label{normc}
\end{equation}
where $I_{i\omega }(x)$ is the Bessel modified function and for a
given function $f(z)$ we use the notation
\begin{equation} \label{barnot}
\bar{f}(z)\equiv
Af(z)+zf^{\prime }(z)
\end{equation}
(not to be confused with the bared
notations in the conformally transformed frame in Sec.
\ref{sec:emtbs}).

Substituting the eigenfunctions (\ref{sol2}) into the mode-sum
formula (\ref{emtvev1}) with the surface energy-momentum tensor
and integrating over the angular part of ${\mathbf{k}}$, for the
corresponding vacuum expectation value one finds
\begin{eqnarray}\label{tauvev}
\langle 0 |\tau _l^k |0\rangle &=& \frac{2\zeta a^{-D}}{(4\pi
)^{\frac{D-1}{2}}\Gamma \left( \frac{D-1}{2} \right) } \left[
\delta _l^0 \delta _0^k+\left( 4\zeta -1\right) (1+m_s a/\zeta
)\delta _l^k \right]  \nonumber \\
&& \times \int_{0}^{\infty }dx x^{D-2}\sum_{n=1}^{\infty
}\frac{\bar I_{i\omega _n}(x) K^2_{i\omega _n}(x)}{\frac{\partial
}{\partial \omega }\bar K_{i\omega }(x)| _{\omega =\omega _n}}
,\quad l,k=0,2,\ldots , D ,
\end{eqnarray}
with $\Gamma (x)$ being the gamma function and $\langle 0 |\tau
_1^1 |0\rangle =0$. Expression (\ref{tauvev}) is divergent and
needs some regularization with the subsequent renormalization. One
can introduce a cutoff function and then can apply to the sum over
$n$ the summation formula derived in Ref. \cite{Saha02} from the
generalized Abel-Plana formula. An alternative approach is to
insert in Eq. (\ref{tauvev}) a factor $\omega _n^{-s}$, to
evaluate the resulting expression in the domain of the complex $s$
plane where it is convergent, and then the result is analytically
continued to the value $s=0$. This approach corresponds to the
zeta function regularization procedure (see, for instance,
\cite{Blau88,Eliz94} and references therein). The analytic continuation
over $s$ can be done by the method used in Ref. \cite{Saha03b} for
the evaluation of total Casimir energy. The corresponding results
will be reported elsewhere. Note that on the plate $\xi =a$,
energy-momentum tensor (\ref{tauvev}) is of perfect fluid type
with the equation of state
\begin{equation}\label{eqstate}
\varepsilon ^{{\mathrm{(surf)}}}=-\left[ 1+\frac{\zeta }{(4\zeta
-1)(\zeta +m_s a)}\right] \cal{P}^{{\mathrm{(surf)}}},
\end{equation}
and effective pressure defined by $\langle 0 |\tau _l^k |0\rangle
=-\delta _l^k \cal{P}^{{\mathrm{(surf)}}}$, $l,k=2,\ldots , D$.
For a minimally coupled scalar field this corresponds to a
cosmological constant induced on the plate.

\section{Surface energy on a spherical shell} \label{sec:sphen}

As an application of the general prescription described in previous section, let us consider a spherical surface with radius $a$ as the boundary $\partial \Sigma $ and the interior region as the manifold $M$ in the case $D=3$. As the vacuum expectation values for the volume part of the energy-momentum tensor and the total Casimir energy in this problem are well-investigated in literature (see, for instance, Refs. \cite{Bord01,Lese96,Cogn01,Saha01}), here we will concentarte on the surface energy. Now in boundary condition (\ref{boundcondquadVs}) one has $K=-2/a$, $n^l=-\delta ^l_1$ and, hence,
\begin{equation}
\left( A_s+a\frac{\partial }{\partial r}\right) \varphi =0,\quad A_s=4\zeta -2m_s a,\quad r=a . \label{sphbc}
\end{equation}
Inside the sphere the complete set of solutions to the field equation, regular at the origin, has the form
\begin{equation}
\varphi _{\alpha }(x)=\frac{\beta _{\alpha }}{\sqrt{r}}J_{\nu }(\lambda r )Y_{ln}(\theta ,\phi )e^{-i\omega t},\quad -l\leq n\leq l,\quad l=0,1,2,\ldots ,\quad \nu =l+\frac{1}{2}, \label{spheigfunc}
\end{equation}
where $\lambda =\sqrt{\omega ^2-m^2}$, $J_{\nu }(z)$ is the Bessel function, and $Y_{lj}(\theta ,\phi )$ are the spherical harmonics. The coefficient $\beta _{\alpha }$ can be found from the normalization condition and is equal to
\begin{equation}
\beta _{\alpha }^2=\frac{\lambda T_{\nu }(\lambda a)}{\omega a },\quad T_{\nu }(z)=\frac{z}{(z^2-\nu ^2)J_{\nu }^2(z)+z^2J_{\nu }^{'2}(z)}. \label{sphbet}
\end{equation}
From boundary condition (\ref{sphbc}) for eigenfunctions (\ref{spheigfunc}) one can see that the possible values for $\lambda $ have to be solutions to the following equation
\begin{equation}
AJ_{\nu }(z)+zJ_{\nu }'(z)=0,\quad z=\lambda a,\quad A=A_s-\frac{1}{2}. \label{sphbc1}
\end{equation}
For real $\nu >-1$ all roots of this equation are simple and real, except the case $A<-\nu $ when there are two purely imaginary zeros. In the following consideration we will assume that $A\geq -1/2 $ with all zeros being real. Let us denote by $\lambda _{\nu ,k}$, $k=1,2,\ldots $, the zeros of the function $AJ_{\nu }(z)+zJ_{\nu }'(z)$ in the right half plane, arranged in ascending order. The corresponding eigenfrequencies are defined by the relation $\omega _{\nu ,k}=\sqrt{\lambda _{\nu ,k}^2/a^2+m^2}$. Note that for the massless case the existence of purely imaginary zeros for Eq. (\ref{sphbc1}) would mean that the corresponding quantum state is unstable.

Substituting Eq. (\ref{spheigfunc}) into Eq. (\ref{Esurf}) and using the addition formula for the spherical harmonics, for the surface energy one obtains
\begin{equation}
E^{{\mathrm{(surf)}}}_{{\mathrm{in}}}=\frac{1-4\zeta }{a} A_s\sum_{l=0}^{\infty }\nu \sum _{k=1}^{\infty }\frac{\lambda _{\nu ,k}^2}{(\lambda _{\nu ,k}^2+A^2-\nu ^2)\sqrt{\lambda _{\nu ,k}^2+\mu ^2}},\quad \mu =m a. \label{sphsurfE}
\end{equation}
As it stands, the right hand side of this equation clearly diverges and needs some regularization. We regularize it by defining the function
\begin{equation}
Z_{{\mathrm{in}}}(s)=\sum_{l=0}^{\infty }\nu \zeta _{{{\mathrm{(in)}}}\nu }(s), \label{sphEsurfs}
\end{equation}
where the related partial zeta function is introduced:
\begin{equation}
\zeta _{{{\mathrm{(in)}}}\nu }(s)=\sum _{k=1}^{\infty }\lambda _{\nu ,k}^2\frac{(\lambda _{\nu ,k}^2+\mu ^2)^{-s}}{\lambda _{\nu ,k}^2+A^2-\nu ^2}. \label{sphzetanus}
\end{equation}
(Here and below we use the subscripts in and ex to distinguish the quantities in the interior and exterior regions.) In accordance with the zeta function regularization procedure, the surface energy is obtained by the analytic continuation of $Z_{{\mathrm{in}}}(s)$ to the value $s=1/2$:
\begin{equation}
E_{{\mathrm{in}}}^{{\mathrm{(surf)}}}=\frac{1-4\zeta }{a}A_s \left. Z_{{\mathrm{in}}}(s) \right| _{s=1/2}. \label{EinZ}
\end{equation}
The starting point of our consideration is the representation of the partial zeta function (\ref{sphzetanus}) in an integral form. This is done in Appendix \ref{sec:app1} by making use the generalized Abel-Plana formula (for applications of the generalized Abel-Plana formula see also Refs. \cite{Rome02,Saha01,Rome01,Saha02,Avag02}). The corresponding integral representation for a general case of a massive scalar field is given by formula (\ref{sumformula3}). Below we will concentrate on the massless case for which from (\ref{sumformula3}) one has
\begin{equation}
\zeta _{{{\mathrm{(in)}}}\nu }(s)=\frac{\sin \pi s}{\pi }\int _{0}^{\infty }dx \, x^{1-2s} \frac{I_{\nu }(x)}{\bar I_{\nu }(x)}. \label{sphzm0}
\end{equation}
This integral representation is valid for $1/2<{{\mathrm{Re}}}\,\, s<1$.
For the analytic continuation of the integral on the right to $s=1/2$ we employ the uniform asymptotic expansions of the modified Bessel functions for large values of the order (see, for instance, Ref. \cite{Abra}). From these expansions one has
\begin{equation}
\frac{I_{\nu }(\nu x)}{\bar I_{\nu }(\nu x)}\sim \frac{t}{\nu }\sum_{k=0}^{\infty }\frac{w_k(t)}{\nu ^k},\quad t=\frac{1}{\sqrt{1+x^2}},
\label{Inuratas}
\end{equation}
where the coefficients $w_{k}(t)$ are combinations of the corresponding coefficients in the expansions for $I_{\nu }(\nu x)$ and $I'_{\nu }(\nu x)$. In our consideration the first three coefficients will be sufficient:
\begin{eqnarray}
&& w_0(t)=1,\quad w_1(t)=\left( \frac{1}{2}-A\right) t-\frac{1}{2}t^3,\nonumber \\
&& w_2(t)=\left( \frac{3}{8}-A+A^2\right) t^2 +\left( A-\frac{5}{4}\right) t^4+\frac{7}{8}t^6 . \label{sphw}
\end{eqnarray}
Note that the functions $w_k(t)$ have the structure
\begin{equation}
w_k(t)=\sum_{j=0}^{k}w_{kj} t^{2j+k}, \label{sphw1}
\end{equation}
where the coefficients $w_{kj}$ depend only on $A$. Following the usual procedure applied in the analogous calculations for the total Casimir energy (see, for instance, Refs. \cite{Lese96,Bord96,Bord01}), we subtract and add to the integrand in Eq. (\ref{sphzm0}) $N$ leading terms of the corresponding asymptotic expansion and exactly integrate the asymptotic part. By this way Eq. (\ref{sphzm0}) may be split into the following pieces
\begin{equation}
\zeta _{{{\mathrm{(in)}}}\nu }(s)=\zeta ^{{\mathrm{(as)}}}_{{{\mathrm{(in)}}}\nu }(s)+\zeta ^{(1)}_{{{\mathrm{(in)}}}\nu }(s) ,\label{sphzsplit}
\end{equation}
where
\begin{eqnarray}
\zeta ^{{\mathrm{(as)}}}_{{{\mathrm{(in)}}}\nu }(s) &=& \nu ^{2(1-s)}\frac{\sin \pi s}{\pi }\int _{0}^{\infty }dx\, x^{1-2s} \sum_{k=0}^{N}\frac{tw_k(t)}{\nu ^{k+1}}, \label{sphzas} \\
\zeta ^{(1)}_{{{\mathrm{(in)}}}\nu }(s) &=& \nu ^{2(1-s)} \frac{\sin \pi s}{\pi }\int _{0}^{\infty }dx\, x^{1-2s} \left[ \frac{I_{\nu }(\nu x)}{\bar I_{\nu }(\nu x)}-\sum_{k=0}^{N}\frac{tw_k(t)}{\nu ^{k+1}}\right] . \label{sphz1}
\end{eqnarray}
The number of terms $N$ in these formulae is determined by the condition of convergence for the expression $\sum _{l=0}^{\infty }\nu \zeta ^{(1)}_{{{\mathrm{(in)}}}\nu }(s)$ at $s=1/2$. It can be seen that it is sufficient to take $N\geq 2$. For larger $N$ the convergence is stronger. Now by taking into account Eq. (\ref{sphw1}) and the expression for $t$, the integral in Eq. (\ref{sphzas}) is easily evaluated in terms of gamma function. For the corresponding sum over $l$ this gives:
\begin{equation}
Z^{{\mathrm{(as)}}}_{{\mathrm{in}}}(s)=\sum_{l=0}^{\infty }\nu \zeta ^{{\mathrm{(as)}}}_{{{\mathrm{(in)}}}\nu }(s)=\frac{1}{2\Gamma (s) }\sum _{k=0}^{N}\zeta _{{\mathrm{H}}}\left( 2s+k-2,\frac{1}{2}\right) \sum_{j=0}^{k}w_{kj}\frac{\Gamma \left( j+\frac{k-1}{2}+s\right) }{\Gamma \left( j+\frac{k+1}{2}\right)} , \label{sphsumz}
\end{equation}
where we have introduced Hurwitz zeta function $\zeta _{{\mathrm{H}}}(s,a)$. In the following calculations we will take the minimal value $N=2$. The expression on the right of Eq. (\ref{sphsumz}) has a simple pole at $s=1/2$ coming from the poles of Hurwitz and gamma functions. Laurent-expanding near this pole one receives
\begin{equation}
Z^{{\mathrm{(as)}}}_{{\mathrm{in}}}(s)=\frac{Z^{{\mathrm{(as)}}}_{-1}}{s-1/2}+Z^{{\mathrm{(as)}}}_{0}+{\cal O} (s-1/2) , \label{Zas}
\end{equation}
where
\begin{eqnarray}
Z^{{\mathrm{(as)}}}_{-1} &=& \frac{1}{2 \pi}\left( A^2-\frac{1}{3}A+\frac{1}{20}\right) \label{Zas-1}\\
Z^{{\mathrm{(as)}}}_{0} &=& \frac{1}{\pi }\left[ A^2\left( \gamma +3\ln 2\right) +\frac{A}{3}\left( 1-\gamma -3\ln 2\right) \right. \nonumber \\
&& -\left. \frac{1}{2}\zeta _{{\mathrm{R}}}'(-1)+\frac{3\ln 2 +\gamma -8}{120} \right] \nonumber \\
&=& \frac{1}{\pi }\left( 0.0382 -0.552 A+2.657 A^2 \right) , \label{Zas0}
\end{eqnarray}
with the Euler's constant gamma and Riemann zeta function $\zeta _{{\mathrm{R}}}(s)$.

As regards to Eq. (\ref{sphz1}), for $N\geq 2$ it is finite at $s=1/2$ (the series over $l$ converges as $1/l^2$) and we can directly put this value and evaluate the integral and sum numerically. Now the expression for the surface energy contains pole and finite parts:
\begin{equation}
E^{{\mathrm{(surf)}}}_{{\mathrm{in}}}=E^{{\mathrm{(surf)}}}_{{\mathrm{(in)p}}}+E^{{\mathrm{(surf)}}}_{{\mathrm{(in)f}}} , \label{Esurfverj}
\end{equation}
where for the separate contributions one has
\begin{equation}
E^{{\mathrm{(surf)}}}_{{\mathrm{(in)p}}}=\frac{1-4\zeta }{a}\frac{A_sZ^{{\mathrm{(as)}}}_{-1}}{s-1/2},\quad E^{{\mathrm{(surf)}}}_{{\mathrm{(in)f}}}=\frac{1-4\zeta }{a}A_s\left[ Z^{{\mathrm{(as)}}}_{0}+\sum _{l=0}^{\infty }\nu \zeta _{{\mathrm{(in)}}\nu }^{(1)}(1/2)\right] . \label{Esurfpf}
\end{equation}
It is well-known that the total vacuum energy inside a spherical shell also contains pole and finite contributions \cite{Bord01,Lese96,Cogn01}. In particular, the asymptotic part of the total Casimir energy inside a spherical shell has a structure similar to Eq. (\ref{sphsumz}) and the corresponding pole part comes again from the poles of Hurwitz and gamma functions. Note that taking interior and exterior space together, in odd spatial dimensions the pole parts cancel and a finite result emerges. Now let us see that the similar situation takes place for the surface energy.

\begin{figure}[tbph]
\begin{center}
\epsfig{figure=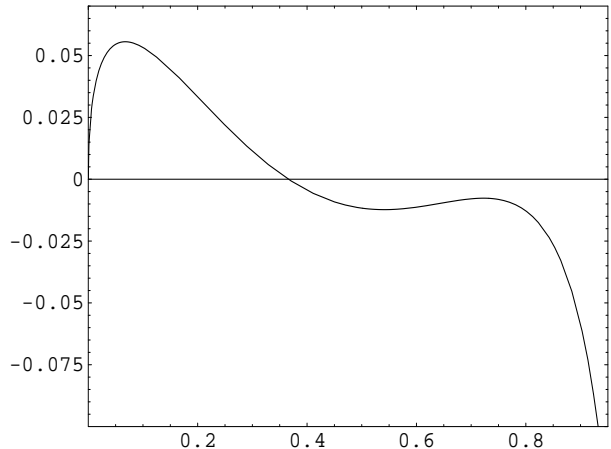,width=8cm,height=6cm}
\end{center}
\caption{ The total surface energy for a spherical shell, $aE^{{\mathrm{(surf)}}}/(1-4\zeta )$, as a function on the coefficient $A_s$ in boundary condition (\ref{sphbc}). }
\label{fig1}
\end{figure}

Consider the scalar vacuum outside a spherical shell on which the field satisfies boundary condition (\ref{sphbc}). To deal with discrete spectrum, we can introduce a larger sphere with radius $R$, enclosing one of radius $a$, on whose surface we impose boundary conditions as well. After the construction of the corresponding partial zeta function we take the limit $R\to \infty $ (for details of this procedure see Ref. \cite{Lese96} in the case of the exterior Casimir energy and Ref. \cite{Saha01} for the corresponding vacuum densities). As a result for the surface energy in the exterior region one obtains
\begin{equation}
E_{{\mathrm{ex}}}^{{\mathrm{(surf)}}}=\frac{1-4\zeta }{a}A_s \left. Z_{{\mathrm{ex}}}(s) \right| _{s=1/2}, \label{EinZext}
\end{equation}
with
\begin{equation} \label{Zext}
Z_{{\mathrm{ex}}}(s)=\sum_{l=0}^{\infty }\nu \zeta _{{{\mathrm{(ex)}}}\nu }(s), \quad \zeta _{{{\mathrm{(ex)}}}\nu }(s)=\frac{\sin \pi s}{\pi }\int _{0}^{\infty }dx \, x^{1-2s} \frac{K_{\nu }(x)}{\bar K_{\nu }(x)}.
\end{equation}
In deriving Eq. (\ref{Zext}) we have assumed the condition $A<1/2$ under which the subintegrand has no poles on the positive real axis. The corresponding expression differs from the interior one by the replacement $I_{\nu }\to K_{\nu }$. This relation between the interior and exterior quantities is well-known for the total Casimir energy and vacuum densities. Now the analytical continuation needed can be done by the way described above for the interior case using the uniform asymptotic expansions for the function $K_\nu (x)$. The corresponding formulae will differ from Eqs. (\ref{sphzas})--(\ref{sphsumz}) by replacements $I_{\nu }\to K_{\nu }$ and $w_k(t)\to (-1)^{k+1}w_k(t)$. Now we see that in the sum
\begin{equation} \label{Zastot}
Z^{{\mathrm{(as)}}}(s) =Z^{{\mathrm{(as)}}}_{{\mathrm{in}}}(s)+Z^{{\mathrm{(as)}}}_{{\mathrm{ex}}}(s),
\end{equation}
the summands with even $k$ and, hence, the poles terms cancel (recall that the pole terms in the separate interior and exterior surface energies are contained in $k=0$ and $k=2$ terms). Moreover, for $s=1/2$ the term with $k=1$ in the expression for $Z^{{\mathrm{(as)}}}(s)$ vanishes as well due to $\zeta _{{\mathrm{H}}}(0,1/2)=0$. In particular, the asymptotic part of the total surface energy vanishes for $N=2$. Taking this value $N$ and using the expression of the function $w_1(t)$, for the total surface energy one obtains
\begin{eqnarray}
E^{{\mathrm{(surf)}}}&=& E^{{\mathrm{(surf)}}}_{{\mathrm{in}}}+E^{{\mathrm{(surf)}}}_{{\mathrm{ex}}} \nonumber \\
&=& \frac{1-4\zeta }{\pi a}A_s\sum _{l=0}^{\infty }\nu \int _{0}^{\infty }dx\, \left[ \frac{I_{\nu }(x)}{\bar I_{\nu }(x)}+\frac{K_{\nu }(x)}{\bar K_{\nu }( x)}-\frac{1}{x^2+\nu ^2}\left( \frac{x^2}{x^2+\nu ^2}-2A\right) \right] , \label{sphentot}
\end{eqnarray}
where we have changed the integration variable $\nu x\to x$.
In Fig. \ref{fig1} we have plotted the quantity $a E^{{\mathrm{(surf)}}}/(1-4\zeta )$ as a function on the Robin coefficient $A_s$. This function vanishes at $A_s\approx 0.37$, has a maximum at $A_s\approx 0.07$ and a local minimum at $A_s\approx 0.54$.

\section{Conclusion} \label{sec:conc}

On background of manifolds with boundaries the physical
quantities, in general, will receive both volume and surface
contributions and the surface terms play an important role in
various branches of physics. In particular, surface counterterms
introduced to renormalize the divergencies in the quasilocal
definitions of the energy for the gravitational field and in
quantum field theory with boundaries are of particular interest.
In the present paper, continuing the investigations started in
Refs. \cite{Rome02,Saha01,Rome01}, we argued that already at
classical level there is a contribution to the energy-momentum
tensor of a scalar field, located on the boundary. We have
considered general bulk and boundary geometries and an arbitrary
value of the curvature coupling parameter for the scalar field.
Our starting point is the action (\ref{action}) with bulk and
surface contributions (\ref{actionB}) and (\ref{actionGH}) (see
also (\ref{corneraction}) for the corner term). The surface action
contains the generalization of the standard Gibbons-Hawking
surface term for the problem under consideration, and an
additional quadratic term. We evaluate the energy-momentum tensor
in accordance with the standard procedure, as the functional
derivative of the action with respect to the metric. In the
variational procedure we held fixed the metric on initial and
final spacelike hypersurfaces $\Sigma _1$ and $\Sigma _2$ (see
Fig. \ref{fig1fold}) and allowed variations on the timelike
boundary $\partial M_s$. The surface energy-momentum tensor is
obtained as the functional derivative with respect to the
variations of the metric tensor on $\partial M_s$. As the various
surface terms arising in the variation procedure are essential for
our consideration, the corresponding calculations are presented in
detail for the bulk and boundary parts of the action in Sec.
\ref{sec:emtbs}. The bulk part of the energy-momentum tensor has
the well-known form (\ref{Tik1}) and the surface energy-momentum
tensor is given by expression (\ref{surfemtnew}) with $\tau _{ik}$
from Eq. (\ref{tauik}), and the delta function locates this part
on the boundary $\partial M_s$. The surface part of the
energy-momentum tensor satisfies orthogonality relation
(\ref{surforth}) and does not depend on the foliation of spacetime
into spacelike hypersurfaces.

Similar to the metric variation, taking the variation of the
action induced by an infinitesimal variation in the field, we have
fixed the field on the initial and final spacelike hypersurfaces
$\Sigma _1$ and $\Sigma _2$ and allowed variations on the
hypersurface $\partial M_s$. From the extremality condition for
the action under the variations of the field on $\partial M_s$ we
have received  the boundary condition on $\partial M_s$, given by
Eq. (\ref{boundcondquadVs}). On the class of solutions to the
field equation, the volume part of the energy-momentum tensor
satisfies the standard covariant conservation equation
(\ref{covdiv0}), and for the surface part we have obtained
equation (\ref{covdivsurf0}) which describes the covariant
conservation of the surface energy-momentum tensor up to the flow
of the energy-momentum across the boundary into the bulk. For a
conformally coupled massless scalar field both volume and surface
energy-momentum tensors are traceless and the boundary condition
for the field on $\partial M_s$ coincides with the conformally
invariant Hawking boundary condition. Further we have considered
the surface energy density for a static manifold with static
boundary and have shown that the expression for this quantity,
previously used in Refs. \cite{Kenn80,Rome02,Saha01,Rome01}, is
obtained as a special case with $g_{00}=1$. As an example of
geometry when the formula used in the papers cited above needs a
generalization, we considered a plate moving with uniform proper
acceleration in the Minkowski bulk. In Sec. \ref{sec:intcons} the
integral conservation laws are investigated which follow from the
covariant differential conservation equations (\ref{covdiv0}) and
(\ref{covdivsurf0}). Under the presence of Killing vectors the
corresponding volume and surface conserved charges are constructed
by formulae (\ref{Pvol}) and (\ref{Psurf}). The total energy
momentum is obtained as a sum of these parts. The difference of
the total energy-momentum for two spacelike hypersurfaces $\Sigma
_1$ and $\Sigma _2$ is related to the projected onto normal $n^i$
part of the volume energy-momentum tensor by Eq.
(\ref{Ptotchange}). The subintegrand on the right of this formula
can be interpreted in terms of the normal force acting on the
constraining boundary. This force is determined by the volume part
of the energy-momentum tensor only. Further, we have shown that
the total energy-momentum can be also obtained by integrating the
total energy-momentum tensor over a spacelike hypersurface $\Sigma
$. We have expressed the volume and surface energy-momenta in
terms of the proper energy volume and surface densities and proper
momentum volume and surface densities, formulae (\ref{Pvoleps})
and (\ref{Psurfeps}). A special case is considered when a timelike
Killing vector exists for the bulk metric. By choosing an
appropriate foliation we have seen that in this case the
corresponding volume energy-momentum coincides with the integrated
volume proper energy. For the general case of the non-orthogonal
foliation the integrated surface energy is given by formula
(\ref{Psurfeps2}). In the case of an orthogonal foliation, which
corresponds to a static boundary, the total energy is a conserved
quantity. We have emphasized that, in general, this is not the
case for separate volume and surface parts.

In Sec. \ref{sec:vacemt} we have considered a massive quantum
scalar field on a static manifold with static boundary. By
integrating the vacuum expectation values of the energy densities
we have found the volume and surface parts of the vacuum energy.
We have shown that (i) the integrated volume energy, in general,
differs from the vacuum energy evaluated as the sum of zero-point
energies for each normal mode and (ii) this difference is due to
the additional contribution coming from the surface part of the
vacuum energy. These results for the special cases of flat,
spherical and cylindrical Robin boundaries on the Minkowski bulk
have been discussed previously in Refs.
\cite{Rome02,Saha01,Rome01}. As an application of general formulae
we have derived a formal expression for the surface
energy-momentum tensor induced on an infinite plane boundary
moving with uniform proper acceleration through the
Fulling-Rindler vacuum. This tensor is of barotropic perfect fluid
type with the equation of state (\ref{eqstate}) and in the case of
a minimally coupled scalar corresponds to a cosmological constant
on the plate.

As an application of a general formula for the surface energy-momentum tensor, in Sec. \ref{sec:sphen} we have evaluated the vacuum expectation value of the surface energy for a quantum scalar field for interior and exterior regions of a spherical shell. As a regularization procedure we have used the zeta function technique. An integral representation for the related partial zeta function well suited for the analytic continuation is constructed in Appendix \ref{sec:app1}, by using the generalized Abel-Plana summation formula over zeros of a combination of the Bessel function with derivatives. Subtracting and adding to the integrands the leading terms of the corresponding uniform asymptotic expansions, we present the interior and exterior surface energies as sums of two parts. The first ones are finite at $s=1/2$ and can be easily evaluated numerically. In the second, asymptotic, parts the pole contributions are given explicitely in terms Hurwitz and gamma functions.
The remained pole terms are characteristic feature of the zeta function regularization method and have been found for many cases of boundary geometries. For an infinitely thin shell
taking interior and exterior regions together, the
pole parts cancel and the total surface energy is finite and
can be directly evaluated by making use formula (\ref{sphentot}). The cancellation of the pole terms coming from
oppositely oriented faces of infinitely thin smooth boundaries
takes place in very many situations encountered in the literature.
It is a simple consequence of the fact that the second fundamental
forms are equal and opposite on the two faces of each boundary
and, consequently, the value of the corresponding coefficient in
the heat kernel expansion summed over the two faces of each
boundary vanishes \cite{Blau88}. In even spatial dimensions there is no
such a cancellation.

In the consideration above we have assumed a smooth timelike
boundary $\partial M_s$. For non-smooth boundaries the corner
terms in the action will be important. In this case, in the
variation of the action with respect to the metric, terms will
appear with the integration over the corners. This terms will lead
to the additional part in the energy-momentum tensor located on
the corners.

\section*{Acknowledgement }

The author acknowledges the hospitality of the Abdus Salam
International Centre for Theoretical Physics, Trieste, Italy. The
work was supported in part by the Armenian Ministry of Education
and Science Grant No. 0887.

\appendix

\section{Integral representation for the partial zeta function} \label{sec:app1}

In this appendix we will derive an integral representation of partial zeta function (\ref{sphzetanus}). For this we use the summation formula over zeros $\lambda _{\nu ,k}$ derived from the generalized Abel-Plana formula \cite{SahaAP,Saha00} (see Eq. (3.26) in Ref. \cite{Saha00}):
\begin{eqnarray}
\sum_{k=1}^{\infty }T_{\nu }(\lambda _{\nu ,k})f(\lambda _{\nu ,k})J_{\nu }^2(\lambda _{\nu ,k}) &=& \frac{1}{2}\int_0^{\infty }dx\, f(x)J_{\nu }^2(x) \nonumber \\
&& - \frac{1}{2\pi }\int_{0}^{\infty }dx\, \frac{\bar K_{\nu }(x)}{\bar I_{\nu }(x)} I_{\nu }^2(x)\left[ f(xe^{\pi i/2})+f(xe^{-\pi i/2})\right] , \label{sumformula}
\end{eqnarray}
where the function $T_{\nu }(z)$ is defined by Eq. (\ref{sphbet}) and we use the notation indroduced in Eq. (\ref{barnot}).
This formula is valid for functions $f(z)$ analytic in the right half plane and satisfying the condition
\begin{equation}
\left| f(z)\right|<\frac{M_0}{|z|^{\alpha }}, \quad \alpha >0, \quad |z|\to \infty , \label{cond1}
\end{equation}
with a constant $M_0$. To transform the first integral on the right of formula (\ref{sumformula}), let us apply this formula to the case $A=0$ (the sum is over zeros of the Bessel function $J_{\nu }(z)$). Now the left hand side is zero and from (\ref{sumformula}) one receives
\begin{equation}
\int_{0}^{\infty }dx\, f(x)J_{\nu }^2(x)=\frac{1}{\pi }\int_{0}^{\infty }dx\, I_{\nu }(x) K_{\nu }(x) \left[ f(xe^{\pi i/2})+f(xe^{-\pi i/2})\right] . \label{intform}
\end{equation}
The resubstitution into Eq. (\ref{sumformula}) yields
\begin{equation}
\sum_{k=1}^{\infty }\frac{\lambda _{\nu ,k}f(\lambda _{\nu ,k})}{\lambda _{\nu ,k}^2+A^2-\nu ^2}=\frac{1}{2\pi }\int_{0}^{\infty }dx\, \frac{I_{\nu }(x)}{\bar I_{\nu }(x)}\left[ f(xe^{\pi i/2})+f(xe^{-\pi i/2})\right] ,\label{sumformula1}
\end{equation}
where we have used the relation $T_{\nu }(\lambda _{\nu ,k}) J_{\nu }^2(\lambda _{\nu ,k})=\lambda _{\nu ,k}/(\lambda _{\nu ,k}^2+A^2-\nu ^2)$ and the Wronskian for the Bessel modified functions. As a function $f(z)$ in this formula let us choose
\begin{equation}
f(z)=\frac{z}{(z^2+\mu ^2)^s} , \label{fz}
\end{equation}
which satisfies the condition (\ref{cond1}) for ${\mathrm{Re}}\, \, s>1/2$. We obtain
\begin{equation}
\sum_{k=1}^{\infty }\frac{\lambda _{\nu ,k}^2(\lambda _{\nu ,k}^2+\mu ^2)^{-s}}{\lambda _{\nu ,k}^2+A^2-\nu ^2}=\frac{\sin \pi s}{\pi }\int_{\mu }^{\infty }dx\, \frac{I_{\nu }(x)}{\bar I_{\nu }(x)} \frac{x}{(x^2-\mu ^2)^{s}} . \label{sumformula3}
\end{equation}
This formula gives an integral representation of the partial zeta function introduced in section~\ref{sec:sphen}.

\end{document}